\documentclass[aps,prb,twocolumn,groupeauthor,showpacs]{revtex4}

\usepackage{amsmath}
\usepackage{graphicx}
\usepackage{dcolumn}
\usepackage{bm}
\usepackage[dvips]{color}
\usepackage{ulem}
\usepackage{multirow}

\begin{document}

\title{Excitation of multiple phonon modes in copper metaborate CuB$_2$O$_4$\\via non-resonant impulsive stimulated Raman scattering}

\author{Kotaro Imasaka,$^1$ Roman V. Pisarev,$^2$ Leonard N. Bezmaternykh,$^3$ Tsutomu Shimura,$^1$ Alexandra M. Kalashnikova,$^{1,2}$\email{kalashnikova@mail.ioffe.ru} and Takuya Satoh$^{1,4}$}
\affiliation{$^1$Institute of Industrial Science, The University of Tokyo, 153-8505 Tokyo, Japan\\
$^2$Ioffe Institute, 194021 St.\ Petersburg, Russia\,\,\,\,\,\,\,\,\,\\
$^3$L. V. Kirensky Institute of Physics, SB RAS, 660036 Krasnoyarsk, Russia\\
$^4$Department of Physics, Kyushu University, 819-0395 Fukuoka, Japan}

\date{\today}

\begin{abstract}
The excitation of four coherent phonon modes of different symmetries are realized in copper metaborate CuB$_2$O$_4$ via impulsive stimulated Raman scattering (ISRS). The phonons are detected by monitoring changes in the linear optical birefringence using the polarimetric detection (PD) technique. We compare the results of the ISRS-PD experiment to the polarized spontaneous Raman scattering spectra. We show that agreement between the two sets of data obtained by these allied techniques in a wide phonon frequency range of 4--14\,THz can be achieved by taking into account the symmetry of the phonon modes and corresponding excitation and detection selection rules. It is also important to account for the difference between incoherent and coherent phonons in terms of their contributions to the Raman scattering process. This comparative analysis highlights the importance of the ratio between the frequency of a particular mode, and the pump and probe spectral widths. We analytically demonstrate that the pump and probe pulse durations of 90 and 50 fs, respectively, used in our experiments limit the highest frequency of the excited and detected coherent phonon modes to 12 THz, and define their relative amplitudes.
\end{abstract}


\maketitle

\section{Introduction}

Impulsive stimulated Raman scattering (ISRS)\cite{deSilvestri-CPL1985} is a powerful technique, allowing the generation of collective excitations in a medium via inelastic scattering of a single (sub-)picosecond laser pulse. It has become a widely used tool for exciting coherent phonons,\cite{Yan-JCP1987,Dhar-ChemRev1994,Merlin-SSC1997,Misochko-JETP2001} magnons,\cite{Kirilyuk-RMP2010,Kalashnikova-PhysUsp2015} phonon-polaritons,\cite{Bakker-RMP1998} and coherent charge fluctuations.\cite{Lorenzana-EPJ2013} ISRS is essentially the stimulated inelastic scattering of a photon of frequency $\omega_i$ into a photon of frequency $\omega_j=\omega_i-\Omega$ accompanied by the creation of a quasiparticle of frequency $\Omega$. This process is, on the one hand, governed by selection rules, i.e., by medium properties described by a Raman tensor, and, on the other hand, highly responsive to the spectral and temporal characteristic of the exciting laser pulse. As a result, ISRS enables selective excitation of particular coherent quasiparticles achieved by choosing proper polarization of the laser pulse. Control of the coherent quasiparticle amplitude can be realized by pulse shaping.\cite{Weiner-Science1990,Weiner-JOSAB1991,Kawashima-ARPC1995,Feurer-Science2003,McGrane-NJP2009,Shimada-APL2012} Importantly, ISRS can be realized in both opaque and transparent media. While in the former case ISRS competes with other excitation mechanisms based on impulsive light absorption,\cite{Zeiger-PRB1992} in the latter case, non-resonant ISRS is the sole mechanism driving coherent excitations of electrons, lattice, or spins.\cite{Bossini-PRB2014}

ISRS and spontaneous Raman scattering (RS) are closely related processes. The selection rules for RS define the polarization of exciting femtosecond pulses to be used for triggering impulsively specific excitation in a medium. However, since the (sub-)picosecond pulse should possess a sufficiently broad spectrum to excite a coherent mode of a given frequency, the pulse duration defines the efficiency of the ISRS. In contrast to RS experiments, ISRS employs probe pulses to monitor excited coherent quasiparticles in the time domain. Therefore, the polarization and duration of these pulses, although often disregarded, are of no lesser importance for the outcome of the conventional ISRS experiments. The ISRS and RS processes differ in the character of the quasiparticles addressed; coherent versus incoherent, respectively. Understanding how the interplay of all these factors affects the results of the ISRS experiment, and establishing vivid links between Raman tensor components and values measured in the ISRS experiments are crucial in light of recent developments of ISRS-based techniques. Thus, a novel approach was recently suggested to obtain information about Raman tensors from ISRS-based coherent lattice fluctuation spectroscopy.\cite{Mann-PRB2015} It has been shown that a comparison of RS and pump-probe data can be used to identify the processes underlying coherent phonon-plasmon mode generation in doped GaN.\cite{Ishioka-JPCM2013} The comparison between RS data and the outcome of pump-probe experiments was also recently made for the case of displacive excitation of coherent phonons (DECP) in opaque bismuth and antimony to obtain insights into ultrafast processes triggered by femtosecond laser pulses.\cite{Li-PRL2013} It was recently suggested in Ref.\,\onlinecite{Nakamura-PRB2015} and later disputed in Ref.\,\onlinecite{Misochko-JETP2016} that the changes in the coherent phonon amplitude with pump pulse duration may shed light on the excitation mechanism and help distinguish resonant ISRS from DECP mechanisms. Finally, the excitation and detection of a plethora of coherent quasi-particles in a single experiment, as well as access to other types of high-frequency collective excitations,\cite{Bossini-NComm2015} has recently become possible owing to the availability of laser pulses of ever-shorter durations.\cite{Brida-JOpt2010} Therefore, tuning polarization,\cite{Satoh-NPhot2015} spectral,\cite{Bossini-PRB2014} temporal,\cite{Shimada-APL2012} and phase \cite{Bardeen-PRL1995,Wand-PCCP2010} characteristics of laser pulses are actively exploited nowadays for realizing selective excitation of particular collective modes.

In this Article we demonstrate how an intrinsic interconnection between the values measured by RS and ISRS techniques can be achieved by designing a \textit{single} ISRS experiment. This reveals the roles of the pump and probe polarizations and durations, as well as the incoherent and coherent natures of involved quasiparticles. All prerequisites for such an experiment are met by impulsively exciting in a dielectric copper metaborate CuB$_2$O$_4$ multiple coherent optical phonon modes of different symmetries, and probing them via a polarization-sensitive optical effect. The choice of CuB$_2$O$_4$ is motivated by its crystallographic structure\cite{Martinez-R-Acta1971} yielding an exceptionally rich phonon spectrum, unique optical,\cite{Pisarev-PRB2010} magnetic,\cite{Roessli-PRL2001} and magneto-optical properties.\cite{Saito-PRL2008,Boldyrev-PRL2015} We demonstrate that laser pulses of 90-fs duration can effectively excite, in CuB$_2$O$_4$, at least four optical phonon modes of $A_1$ and $B_1$ symmetries with frequencies between 4 and 12 THz. These are detected in the ISRS experiment with the polarimetric detection (PD) technique, in which the polarization modulation of the 50-fs probe pulses is monitored. By comparing the results of our experiment to the RS spectra,\cite{Pisarev-PRB2013,Ivanov-PRB2013} we establish the link between the values measured by these two complimentary techniques, and show that an analysis of the efficiencies of excitation of multiple modes via ISRS has to be performed by taking into account the symmetry of each mode, the corresponding excitation and detection selection rules, and the ratio between the frequency of the particular mode and the pump and probe spectral widths.
We note that our approach of revealing the role of the pump and probe pulse durations in ISRS is an alternative to the conventional one, when one tunes the duration of the pump or probe pulses and monitors the corresponding changes in a particular excited coherent phonon mode in a medium.\cite{Yee-JKPS2003,Misochko-JETP2016} In latter studies special care had to be taken to account for the positive or negative chirp of either pump or probe pulses, having different effects on the amplitudes of excited\cite{Bardeen-PRL1995,Bardeen-JCPA1998,Misochko-APL2007} and detected\cite{Monacelli-JCPL2017} coherent quasiparticles.

This Article is organized as follows. In Sec.~\ref{Sec-CuBO} we briefly discuss the copper metaborate properties. In Sec.~\ref{Sec-exp} we describe the sample of copper metaborate CuB$_2$O$_4$ and the details of the ISRS-PD experiment. In Sec.~\ref{Sec-resultsExp} we present the experimental data of the excitation and detection of multiple coherent phonons in CuB$_2$O$_4$ by femtosecond laser pulses. In Sec.~\ref{Sec-discussion} we introduce the formalism for describing the ISRS excitation and polarimetric detection of coherent phonons. In Sec.~\ref{Sec-polarization} we compare the outcomes of the ISRS-PD experiments with the spontaneous RS spectra and analyze the excitation mechanism and specific detection features of the techniques employed. This is followed by an analysis and discussion in Sec.~\ref{Sec-duration} regarding the effect of the pump and probe pulses durations on the excitation and detection of coherent phonons. In Sec.~\ref{Sec-conclusion} we summarize our findings and discuss their eventual impact on further studies of ultrafast laser-induced processes.

\section{Crystal structure and lattice excitations in a copper metaborate C\lowercase{u}B$_2$O$_4$}\label{Sec-CuBO}

CuB$_2$O$_4$ crystallizes in the tetragonal non-centrosymmetric space group \textit{I}$\bar{4}$2\textit{d}, and its primitive unit cell contains 42 atoms.\cite{Martinez-R-Acta1971} This results in 126 zone-center phonon modes, including three acoustical and 123 optical. Jahn-Teller Cu$^{2+}$ (3$d^9$) ions occupy two non-equivalent crystallographic positions, $8d$ and $4b$, in a strongly elongated [Cu$^{2+}$O$^{2-}_6$] octahedron and planar [Cu$^{2+}$O$^{2-}_4$] complex, respectively. This unique structure yields nontrivial optical,\cite{Pisarev-PRB2010} phonon,\cite{Pisarev-PRB2013,Ivanov-PRB2013,Tomov-JPhysCS2016} and magnon,\cite{Boehm-JMMM2002,Martynov-JMMM2006,Ivanov-PRB2013} spectra of this compound.

The fundamental optical band gap of copper metaborate is $\sim$4 eV.\cite{Pisarev-PRB2010} The polarized optical absorption spectra below the fundamental band gap are characterized by an exceptionally pronounced set of zero-phonon lines arising from $3d-3d$ localized electronic transitions in Cu$^{2+}$ ions in two positions, and accompanied by multiple phonon-assisted sidebands.\cite{Pisarev-PRB2010} This observation has naturally triggered an interest in experimental and theoretical analyses of the phonon modes in CuB$_2$O$_4$ by means of infrared and Raman spectroscopy in a wide temperature range of 4--300 K.\cite{Pisarev-PRB2013,Ivanov-PRB2013,Tomov-JPhysCS2016} All theoretically predicted optical phonon modes in the center of the Brillouin zone were observed in the frequency range above 4 THz and assigned to particular atomic motions.\cite{Pisarev-PRB2013} Below $\sim$15 THz phonon spectra are dominated by vibrations in Cu-O complexes, while the dynamics of B-O complexes contributes to the higher-frequency phonon modes. Some of these modes involve vibrations within [Cu$^{2+}$O$^{2-}_4$] exclusively, while no such pure modes exists for the other complex.

It is worth noting that an intricate magnetic structure of CuB$_2$O$_4$ is described by two magnetic sublattices comprised by Cu$^{2+}$ ions in the $8d$ and $4b$ positions. Strong exchange interactions were found only within the sublattice formed by Cu$^{2+}(4b)$ magnetic moments below the N\'{e}el temperature, $T_N$=21 K. Intersublattice interactions yield an ordering of the second sublattice below $\approx$10 K.\cite{Boehm-JMMM2002,Martynov-JMMM2006} As a result, CuB$_2$O$_4$ possesses a very rich magnetic phase diagram,\cite{Roessli-PRL2001,Boehm-PRB2003,Pankrats-JETPLett2003,Fiebig-JAP2003,Petrova-JETP2018} the details of which were clarified experimentally only recently and are not yet fully understood.\cite{Boldyrev-PRL2015} Driving coherent phonon modes, in particular pure [Cu$^{2+}$O$^{2-}_4$] modes, can be seen as a prospective approach for exploiting phonon-magnon interactions in such a complex system.

In the last few years, there have been a number of intriguing and controversial reports of the magneto-optical properties of copper metaborate,\cite{Saito-PRL2008,Lovesey-JPCM2009-1,Arima-JPCM2009,Lovesey-JPCM2009-2,Boldyrev-PRL2015,Toyoda-PRL2015,Lovesey-PRB2016,Nii-JPSJ2017} including the very recent demonstration of laser-induced non-reciprocal light absorption.\cite{Bossini-NPhys2018}

\section{Experimental}\label{Sec-exp}

\begin{figure}
\includegraphics[width=7.6cm]{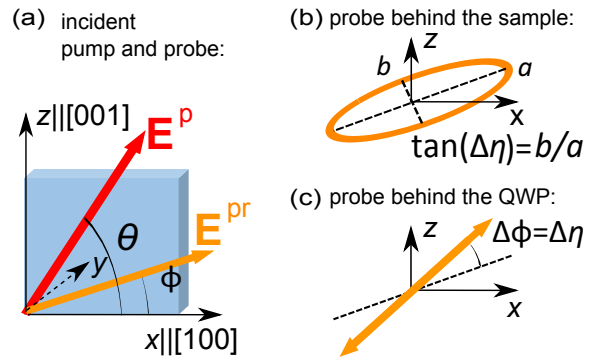}
\caption{(Color online) (a) Relative orientation of the crystallographic axes and laboratory frame $xyz$. The pump pulses are incident along the $y$-axis. The angle of incidence of the probe beam is 7$^\circ$. The pump ($\mathbf{E}^\mathrm{p}$) and probe ($\mathbf{E}^\mathrm{pr}$) pulses are linearly polarized with the azimuthal angles $\theta$ and $\phi$, respectively. (b) Elliptically polarized probe pulse after transmission through the sample. (c) Ellipticity, $\Delta\eta$, of the probe pulses converted to a rotation $\Delta\phi$ of the polarization plane after transmission through the quarter-wave plate.}
\label{Fig:experimental}
\end{figure}

The sample was a plane-parallel single crystal plate cut perpendicular to the [010] axis from a large boule grown by the Kyropulos method from the melt of the oxides B$_2$O$_3$, CuO, Li$_2$O, and MoO$_3$.\cite{ALeksandrov-PSS2003} We used a coordinate system with the $x$-, $y$-, and $z$-axes directed along the [100], [010], and [001] crystallographic axes, respectively [Fig.~\ref{Fig:experimental}(a)]. We note that the [100] and [010] axes cannot be distinguished in the paramagnetic phase, and the assignment of $x$ and $y$ to these axes was performed for the sake of convenience. The thickness of the sample was $d_0=$67 $\mu$m.

The experimental studies of the excitation of the coherent phonons by femtosecond laser pulses were performed using an optical pump-probe technique. The 50-fs laser pulses with a central photon energy of $1.55$ eV at a repetition rate of 1\,kHz were generated by a regenerative amplifier. Here and below we define the pulse duration as the full width at the half maximum (FWHM) of its intensity profile. Part of the output beam was steered to the optical parametric amplifier, producing $\tau_\mathrm{p}=90$-fs pump pulses of $\hbar\omega_\mathrm{p}=1.08$-eV central photon energy. CuB$_2$O$_4$ is transparent in this spectral range, and so ISRS was the dominant mechanism for coherent phonon excitation. Linearly polarized pump pulses at the azimuthal angle $\theta$ with respect to the $x$-axis [see Fig.~\ref{Fig:experimental}(a)] propagated along the sample normal. The pump spot size at the sample was 70 $\mu$m (FWHM). A second part of the regenerative amplifier output beam was used as the probe pulses ($\hbar\omega_\mathrm{pr}=1.55$ eV, $\tau_\mathrm{pr}=50$ fs) and was delayed with the respect to the pump pulses by the variable time,  $t$. The incident probe pulses were linearly polarized with the azimuthal angle $\phi$ with respect to the $x$-axis [Fig.~\ref{Fig:experimental}(a)]. The probe spot size at the sample was 40 $\mu$m (FWHM).

Coherent phonons excited by the pump pulses modulate the dielectric tensor components of CuB$_2$O$_4$, which can be seen in the experiment as a modulation either of the probe pulse ellipticity [Fig.~\ref{Fig:experimental}(b)] or the probe pulse polarization azimuthal angle due to the pump-induced changes in the crystallographic linear birefringence or dichroism, respectively. We note that the absorption coefficient of CuB$_2$O$_4$ at the probe photon energy is $\sim$60 cm$^{-1}$, which suggests that the dichroism experienced by the probe pulses is relatively weak. In the experiments we measured the pump-induced probe ellipticity changes $\Delta\eta$ by employing the PD technique. A quarter-wave plate (QWP) placed in the probe beam behind the sample was used to convert the ellipticity $\Delta\eta$ to a polarization rotation $\Delta\phi$ [Fig.~\ref{Fig:experimental}(c)]. The probe beam transmitted through the sample and QWP was split by a Wollaston prism into vertically and horizontally polarized beams and their intensities were detected by two Si-photodiodes. In this way, the change of the polarization of the probe pulses, measured as the difference between the signals at the two diodes, was monitored as a function of the pump-probe time-delay $t$. The pump-probe traces were recorded in steps of 0.02\,ps up to 10\,ps, including a $-2$\,ps negative pump-probe delay. All measurements were performed at $T=293$ K.

We would like to emphasize that the detection of coherent phonons, excited via ISRS, is usually realized via the monitoring of changes in the reflectivity or transmitivity. Instead, we employed the PD scheme to reveal transient polarization changes, which is more common for experiments on coherent magnons excited via ISRS.\cite{Kalashnikova-PhysUsp2015} To emphasize the differences with the conventional ISRS experiments with coherent phonons, further on we refer to our experimental layout as to ISRS-PD experiment. An advantage of such a scheme for coherent phonon detection is that it allows the analysis of the symmetry of particular phonon modes and can discriminate between them if necessary by choosing a proper probe polarization, as we discuss in detail below. We note that such a scheme is a powerful alternative to the reflective electro-optical sampling technique.\cite{Kutt-IEEE1992,Min-APL1990} The polarization sensitive detection of coherent phonons was reported in e.g., Ref.\,\onlinecite{Dekorsy-PRL1995}, where the polarization dependent reflectivity was analyzed. Here we employ measurements of the transient birefringence, which is an advantageous technique for transparent media.

\section{Excitation and detection of coherent phonons in C\lowercase{u}B$_2$O$_4$}\label{Sec-resultsExp}

\begin{figure}
\includegraphics[width=8.3cm]{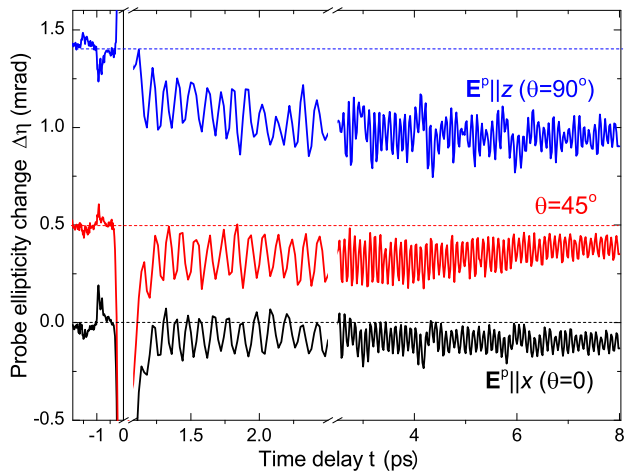}
\caption{(Color online) Probe ellipticity change $\Delta\eta$ as a function of the pump-probe time delay, $t$, measured at three different polarizations, $\theta$, of the pump pulses. The probe polarization is $\phi=45^\circ$.}
\label{Fig:AC_pump}
\end{figure}

\begin{figure*}[t]
\includegraphics[width=16cm]{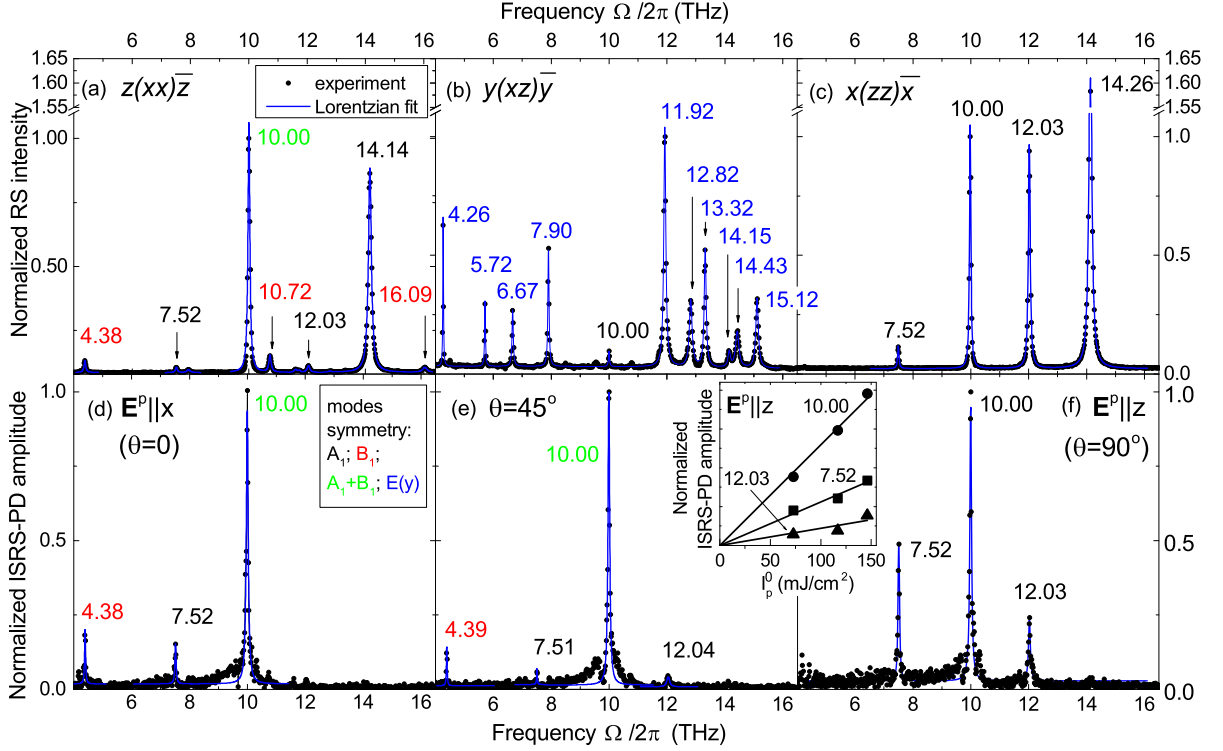}
\caption{(Color online) (a--c) Spontaneous RS spectra measured in the geometries (a) $z(xx)\bar{z}$, (b) $y(xz)\bar{y}$, and (c) $x(zz)\bar{x}$ (adapted from Ref.~\onlinecite{Pisarev-PRB2013}). (d--f) FFT spectra of the ISRS-PD experimental data (Fig.~\ref{Fig:AC_pump}) measured in the geometries (d) $\mathbf{E}^\mathrm{p}\|x$ ($\theta=0$), (e) $\theta=45^\circ$, and (f) $\mathbf{E}^\mathrm{p}\|z$ ($\theta=90^\circ$). In the data sets (a,c) the spectra were normalized by the intensity of the 10.00-THz line, while in the data set (b) -- by the intensity of the strong 11.92-THz line. In the data sets (d--f) the spectra were normalized by the amplitude of the 10.00-THz line. The symbols represent the experimental data, and the lines show their fit using Lorentzian functions [Eq.\,(\ref{eq:Lorentz})]. The numbers indicate the phonon frequencies $\Omega_k/2\pi$. The amplitudes $C_k$ and FWHMs $\sigma_k$ of the phonon lines are given in Tables \ref{Table:PhononsA1B1} and \ref{Table:PhononsA1}. Different colors indicate the phonon modes of the $A_1$ (black), $B_1$ (red), and $E(y)$ (blue) symmetries. The green color denotes the lines to which the phonons of both $A_1$ and $B_1$ symmetries contribute. The inset shows the pump fluence dependence on the amplitudes of the 7.52- (squares), 10.00- (circles), and 12.03-THz (triangles) lines in the FFT spectra shown in panel (f). The lines are linear fits.}
\label{Fig:RS}
\end{figure*}

Figure \ref{Fig:AC_pump} shows the time-delay dependence of the probe polarization excited by the pump pulses of three different polarizations, $\theta=0^\circ$ ($\mathbf{E}^\mathrm{p}\|x$), $\theta=90^\circ$ ($\mathbf{E}^\mathrm{p}\|z$), and $\theta=45^\circ$. The probe was polarized at $\phi=45^\circ$. In all three excitation geometries, a strong coherent artifact was observed at the pump-probe overlap ($t=0$), followed by a pronounced oscillatory signal, consisting of several superimposed harmonic components. The fast Fourier transforms (FFT) amplitude spectra of the traces at positive delays $t>1$ ps are plotted in Figs.~\ref{Fig:RS}(d--f). In the FFT spectra, four lines with frequencies of 4.38, 7.52, 10.00, and 12.03 THz are clearly distinguished.

To reveal the nature of the observed oscillations of the probe polarization and to examine their relation to the coherent phonons excited by the pump pulses, we compared the FFT spectra (Fig.~\ref{Fig:RS}(d--f)) to the spontaneous RS spectra at 293~K reported in Ref.~\onlinecite{Pisarev-PRB2013} [Figs.~\ref{Fig:RS}(a--c)]. The frequencies of the lines appearing in the FFT spectra in our experiments at $\theta=0,90^\circ$ [Figs.~\ref{Fig:RS}(d,f)] are in excellent agreement with the lowest phonon lines in the RS spectra [Figs.~\ref{Fig:RS}(a,c)]. In particular, the phonons appearing in the $x(zz)\bar{x}$ RS spectrum [Fig.~\ref{Fig:RS}(c)] are excited by the pump pulses polarized along the $z$-axis and propagating along the $y$-axis [Figs.~\ref{Fig:RS}(f)]. We note that such a comparison is justified because the [100] ($x$) and [010] ($y$) crystallographic axes in CuB$_2$O$_4$ are equivalent. Analogously, there is a correspondence between the spontaneous $y(xx)\bar{y}$ RS spectra\cite{Pisarev-PRB2013,Ivanov-PRB2013} and our data obtained for the pump pulses polarized along the $x$-axis. Such a good agreement allows us to assign the observed oscillations of the probe polarization to the modulation of the dielectric permittivity by coherent phonons excited by the pump pulses.

From the comparison of our experimental data with the results of the spontaneous RS experiments, we can determine the particular coherent phonon modes that were excited in each geometry. Pump pulses polarized along the $z$-axis excite non-polar $A_1$ modes with frequencies of 7.52, 10.00, and 12.03 THz, while the pump pulses polarized along the $x$-axis excite the 4.38- and 10.00-THz non-polar $B_1$ modes in addition to the $A_1$ modes (7.52 and 10.00 THz).

For the geometry with the pump pulses polarized at an angle $\theta=45^\circ$, the excitation of the modes with $E(y)$ symmetry are expected, which appear in the RS spectrum\cite{Pisarev-PRB2013} measured in the $y(xz)\bar{y}$ configuration [Fig.~\ref{Fig:RS}(b)]. However, this is not the case, as can be seen in Fig.~\ref{Fig:RS}(e).

In Fig.~\ref{Fig:probe_dep}(a) we show the pump-probe traces obtained for the pump polarization $\theta=90^\circ$ for three distinct probe polarizations; $\phi=45^\circ$, $90^\circ$, and $135^\circ$. In Fig.~\ref{Fig:probe_dep}(b) we plot the amplitude of the oscillatory signals $\Delta\eta_0$ versus the incident probe polarization azimuthal angle $\phi$ at a frequency of $\Omega_0/2\pi=10.00$~THz (the most pronounced oscillatory contribution to the signal [Fig.~\ref{Fig:RS}(f)]).
The amplitudes, $\Delta\eta_0$, were extracted from the fit of the experimental data to the sine-function. We note that here and elsewhere in the text $\Delta\eta_0$ is defined as the signed amplitude of the measured signal. The oscillation amplitude is strongly enhanced when the incoming polarization of the probe pulse makes an angle of $\phi=45^\circ$ with the $x$-axis, thus demonstrating the crucial role of the probe pulse polarization in the ISRS-PD experiment. In particular, the results in Figs.~\ref{Fig:probe_dep}(a,b) suggest that the probe polarized at a nonzero angle with respect to the pump polarization plane favors detection of the coherent phonon mode excited by the latter.\cite{Satoh-NPhot2015} This provides a hint for explaining the absence of the coherent phonon modes in the $E(y)$ symmetry for the signal [Figs.\,\ref{Fig:AC_pump} and \ref{Fig:RS}(e)] measured with $\theta=45^\circ$ and $\phi=45^\circ$.

The most evident difference between the results of the ISRS-PD experiments and spontaneous RS data is that the relative values of the amplitudes of the coherent phonons excited via ISRS cannot be directly related to the amplitudes of the corresponding lines in the spontaneous RS spectra. Furthermore, no coherent phonons with frequencies above 12.03 THz could be reliably observed in the ISRS-PD experiments, despite the fact that some of these yielded very strong lines in the RS spectra (Fig.~\ref{Fig:RS}). In Sec.~\ref{Sec-discussion}--\ref{Sec-duration} we consider the excitation and detection of coherent phonons in detail to account for these observations.

\section{Theoretical background and discussion of ISRS-PD in C\lowercase{u}B$_2$O$_4$}\label{Sec-results}

\subsection{ISRS as the excitation mechanism of coherent phonons in CuB$_2$O$_4$}\label{Sec-discussion}

In general, there are two mechanisms, ISRS and DECP, which can mediate the excitation of coherent phonons by the femtosecond laser pulse. We argue that, in our experiments, ISRS is the mechanism responsible for the excitation. First, the sample is transparent for the pump central photon energy, which suppresses the alternative DECP mechanism.\cite{Kutt-IEEE1992} Furthermore, DECP is expected to only drive symmetric $A_1$ modes.\cite{Zeiger-PRB1992} Indeed, when the DECP mechanism is involved, the ions are driven from the equilibrium positions at the ground state to the non-equilibrium positions at the excited state. As a result, the symmetry of the crystal remains unchanged and only the symmetric vibrational $A_1$ mode can be excited.

\begin{figure}
\includegraphics[width=9cm]{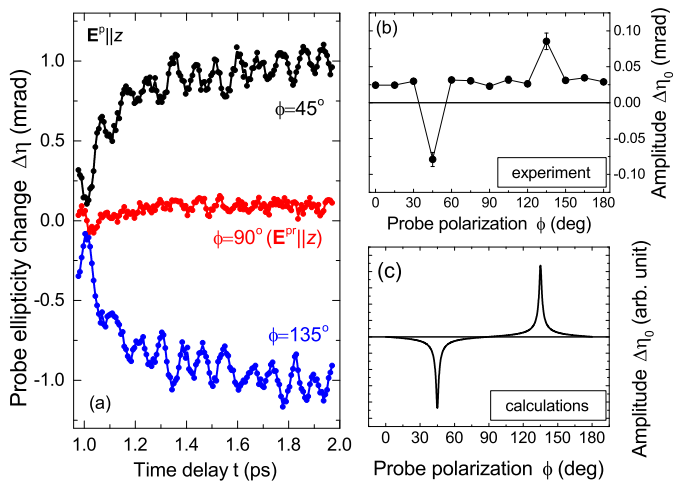}
\caption{(Color online) (a) Probing ellipticity, $\Delta\eta$, versus the pump-probe time delay, $t$, measured for three initial polarizations, $\phi$, of the probe pulses, with a pump polarization of $\theta=90^\circ$. (b) Amplitude of the pump-induced oscillations at a frequency 10.00~THz of the probe polarization $\Delta\eta_0$ as a function of probe azimuthal angle $\phi$. (c) Calculated amplitude of the pump-induced oscillations of the probe polarization as a function of $\phi$ (see the text for details).}
\label{Fig:probe_dep}
\end{figure}

The ISRS-driven excitation of a coherent phonon with frequency $\Omega_0$ and normal coordinate $Q(t)$ can be phenomenologically described by the equation of motion:\cite{Merlin-SSC1997,Satoh-NPhot2015}
\begin{eqnarray}
  \frac{d^2 Q}{dt^2}&+&\Omega_0^2 Q=\nonumber\\
  &&\frac{1}{16\pi}\mathcal{R}^{\Omega_0}_{ij}(\omega_\mathrm{p})\mathrm{Re}[\mathcal{E}_i(t)\mathcal{E}_j(t)^*]=\nonumber\\
  &&\frac{1}{4nc}\mathcal{R}^{\Omega_0}_{ij}(\omega_\mathrm{p})I^0_\mathrm{p}\mathrm{Re}[e_ie_j^*]\alpha(\tau_\mathrm{p},\Omega_0),\label{ISRS_ke}
\end{eqnarray}
where $\mathcal{R}^{\Omega_0}_{ij}(\omega_\mathrm{p})=\partial \varepsilon_{ij}(\omega_\mathrm{p})/\partial Q$ is the Raman tensor and $\varepsilon_{ij}(\omega_\mathrm{p})$ is the dielectric permittivity tensor. Note that Eq.~(\ref{ISRS_ke}) does not include energy dissipation. It is also assumed that the dielectric permittivity dispersion is negligible within the spectral width of the pump pulse. $\mathcal{E}_i(t)$ is the time-dependent electric field envelope defined as $E_i (t) \equiv \mathrm{Re}[\mathcal{E}_i(t) \mathrm{e}^{i \omega_\mathrm{p}t}]$ with amplitude $\mathcal{E}_0$ and FWHM $\tau$. The pump intensity $I^0_\mathrm{p}$ is introduced as $I^0_\mathrm{p}=nc|\mathcal{E}_0|^2/(4\pi)$, where $n$ is the refractive index at the pump wavelength. $\mathbf{e}=(e_x,\,e_y,\,e_z)$ is the polarization unit vector. $\alpha(\tau_\mathrm{p},\Omega_0)$ includes the relation between the duration of the pump pulse, or its spectral width $\sigma_\mathrm{p}$, and the frequency of the coherent phonon mode. The coefficient $\alpha(\tau_\mathrm{p},\Omega_0)$ in Eq.~(\ref{ISRS_ke}) for the phonon mode $\Omega_0$ can be obtained by recalling that the product of two functions in the time domain can be expressed via the convolution of their Fourier-transforms:
\begin{equation}
\alpha(\tau_\mathrm{p},\Omega_0)=\frac{\int_{-\infty}^{\infty}\mathcal{E}_\mathrm{p}(\omega)\mathcal{E}_\mathrm{p}(\omega-\Omega_0)d\omega}{\int_{-\infty}^{\infty}\mathcal{E}_\mathrm{p}(\omega)\mathcal{E}_\mathrm{p}(\omega)d\omega}.
\label{eq:alpha}
\end{equation}
The nominator in the expression for $\alpha(\tau_\mathrm{p},\Omega_0)$ reflects the physical description of the ISRS process, in which pairs of photons with frequencies that differ by $\Omega_0$ contribute to the generation of the corresponding coherent phonon mode. Both the nominator and denominator reduce to $4\pi I^0_p/nc$ for the $\delta-$pulse, and $\alpha(\tau_\mathrm{p}\rightarrow0,\Omega_0)=1$.
The solution of Eq.~(\ref{ISRS_ke}) takes the simple form of a sine-function
\begin{equation}
 Q(t)\sim\alpha(\tau_\mathrm{p},\Omega_0)\mathcal{R}^{\Omega_0}_{ij}(\omega_\mathrm{p})I^0_\mathrm{p}\mathrm{Re}[e_ie_j^*]\sin\left(\Omega_0t\right).\label{eq:NC}
\end{equation}
In the ISRS-PD experiments reported here, the measured value is the change of the ellipticity $\Delta\eta$ of the probe polarization occurring due to modulation of the dielectric permittivity by excited coherent phonons, and converted to the polarization rotation by QWP [Figs.~\ref{Fig:experimental}(b,c)]. The temporal evolution of $\Delta\eta$ can be then expressed as (see App.\,\ref{AppI} for details)
\begin{equation}
\Delta\eta(t)=\frac{\pi d_0}{2\lambda_\mathrm{pr}}\frac{\mathcal{R}^{\Omega_0}_{ij}(\omega_\mathrm{pr})}{\sqrt{\varepsilon_{ij}}}Q(t)\ast \frac{I_\mathrm{pr}(t)}{I^0_\mathrm{pr}},\label{eq:oscillations}
\end{equation}
where $I_\mathrm{pr}(t)$ is the temporal profile of the probe pulse, the exact form of which is introduced below, and $\ast$ denotes the convolution operation. The convolution with the probe pulse temporal profile $I_\mathrm{pr}(t)$ is required to account for the particular probe duration $\tau_\mathrm{pr}$. $\lambda_\mathrm{pr}$ is the probe wavelength. Here, the dielectric permittivity dispersion is assumed negligible within the spectral width of the probe pulse.

The amplitude of the coherent phonons excited and detected in the ISRS-PD process [Eq.\,(\ref{eq:oscillations})] has four main constituents:
\begin{itemize}
\item[(i)] Specific values of Raman tensors $\mathcal{R}^{\Omega_0}_{ij}(\omega_\mathrm{p})$ and $\mathcal{R}^{\Omega_0}_{ij}(\omega_\mathrm{pr})$ are determined by the material properties at the frequencies $\omega_\mathrm{p}$ and $\omega_\mathrm{pr}$ of the pump and probe pulses, respectively. Generally speaking, the absolute values of the $\mathcal{R}^{\Omega_0}_{ij}$ components at these two optical frequencies can differ due to a dispersion in the corresponding spectral range.
\item[(ii)] The product $\mathcal{R}^{\Omega_0}_{ij}(\omega_\mathrm{p})I^0_\mathrm{p}\mathrm{Re}[e_ie_j^*]$ describes the dependence of the driving force on the intensity $I^0_\mathrm{p}$ and polarization $\mathbf{e}$ of the pump pulses.
\item[(iii)] The parameter $\alpha(\tau_\mathrm{p},\Omega_0)$ describes the role of the limited spectral width of the pump pulse in the excitation process.
\item[(iv)] The convolution with $I_\mathrm{pr}(t)$ allows us to account for the polarization changes of the probe pulse due to the modulation of the dielectric permittivity within the range $\tau_\mathrm{pr}$ near the time delay, $t$.
\end{itemize}

\subsection{Role of the pump and probe polarizations in the ISRS-PD experiment}\label{Sec-polarization}

First, we consider the excitation and detection of coherent phonons in CuB$_2$O$_4$, neglecting the duration of the pump and probe pulses, i.e., setting $\alpha(\tau_p,\Omega_0)=1$ and $I_\mathrm{pr}(t)=I^0_\mathrm{pr}\delta(t)$. In this case Eq.\,(\ref{eq:oscillations}) is simplified and takes a form $\Delta\eta(t)=\Delta\eta_0\sin(\Omega_0t)$, where:
\begin{equation}
\Delta\eta_0=\frac{\pi d_0}{2\lambda_\mathrm{pr}}\frac{\mathcal{R}^{\Omega_0}_{ij}(\omega_\mathrm{p})\mathcal{R}^{\Omega_0}_{ij}(\omega_\mathrm{pr})}{\sqrt{\varepsilon_{ij}}}I^0_\mathrm{p}\mathrm{Re}[e_ie^*_j].\label{eq:oscilltions-simple}
\end{equation}
The Raman tensor components for a mode of particular symmetry in CuB$_2$O$_4$ belonging to the point group $\bar{4}2m$ are listed in Table~\ref{Table:driving}.\cite{Pisarev-PRB2013} The modes of the symmetry $A_2$ are silent. Considering the expressions for the Raman tensor components $\mathcal{R}_{ij}$, we can write the r.h.s.\ of Eq.~(\ref{ISRS_ke}) in the form given in Table~\ref{Table:driving}. From this symmetry analysis one can see directly that the $B_2$ and $E(x)$ modes can be excited under the conditions $e_x e_y^*+e_y e_x^* \neq 0$ and $e_y e_z^*+e_z e_y^* \neq 0$, respectively. This is not the case in our experiments, since $e_y=0$ when the pump pulses propagate along the $y$-axis.

\begin{widetext}
\begin{table*}[htb]
  \caption{Raman tensor components $\mathcal{R}_{ij}$ and the corresponding driving forces in Eq.~(\ref{ISRS_ke}) for the phonon modes of a particular symmetry under the assumption of infinitesimally short pump pulses. Also listed are the dielectric tensor components $\delta\varepsilon_{ij}(t)$ modulated by coherent phonons of the particular symmetry. For convenience we have omitted the factor $I^0_\mathrm{p}/4nc$ in the expressions for the driving forces.}
  \begin{center}
    \begin{tabular}{c|c|c|c}
      \hline \hline
      \begin{tabular}{c}Phonon\\symmetry\end{tabular} & $\mathcal{R}_{ij}$ & ISRS driving force & $\delta\varepsilon_{ij}(t)=\mathcal{R}_{ij}Q(t)$\\ \hline
      $A_1$ & $\mathcal{R}_{xx}=\mathcal{R}_{yy},\,\mathcal{R}_{zz}$ & $\mathcal{R}_{xx}\mathrm{Re}\left(e_x e_x^*+e_y e_y^*\right)+\mathcal{R}_{zz}\mathrm{Re}\left(e_z e_z^*\right)$ & $\delta\varepsilon_{xx}(t)=\delta\varepsilon_{yy}(t),\delta\varepsilon_{zz}(t)$\\
      $B_1$ & $\mathcal{R}_{xx}=-\mathcal{R}_{yy}$ & $\mathcal{R}_{xx}\mathrm{Re}\left(e_x e_x^*-e_y e_y^*\right)$ & $\delta\varepsilon_{xx}(t)=-\delta\varepsilon_{yy}(t)$\\
      $B_2$ & $\mathcal{R}_{xy}=\mathcal{R}_{yx}$ & $\mathcal{R}_{xy}\mathrm{Re}\left(e_x e_y^*+e_y e_x^*\right)$ & $\delta\varepsilon_{xy}(t)=\delta\varepsilon_{yx}(t)$\\
      $E(x)$ & $\mathcal{R}_{yz}=\mathcal{R}_{zy}$ & $\mathcal{R}_{yz}\mathrm{Re}\left(e_y e_z^*+e_z e_y^*\right)$ & $\delta\varepsilon_{yz}(t)=\delta\varepsilon_{zy}(t)$\\
      $E(y)$ & $\mathcal{R}_{zx}=\mathcal{R}_{xz}$ & $\mathcal{R}_{xz}\mathrm{Re}\left(e_z e_x^*+e_x e_z^*\right)$ & $\delta\varepsilon_{zx}(t)=\delta\varepsilon_{xz}(t)$\\
      \hline \hline
    \end{tabular}
  \end{center}
   \label{Table:driving}
\end{table*}
\end{widetext}

The driving force for the non-polar $A_1$ modes can be nonzero for any polarization. In the considered geometry the pump pulse of any linear polarization excites these modes, provided the Raman tensor components $\mathcal{R}_{xx}$ and $\mathcal{R}_{zz}$ are nonzero for that particular mode. Indeed, this is observed in our experiment, where the three lowest $A_1$ modes at 7.52-, 10.00-, and 12.03~THz are all excited by the pump pulse polarized along the $z$-axis [Fig.~\ref{Fig:RS}(f)], because all relevant Raman tensor components $\mathcal{R}_{zz}$ are nonzero [Fig.~\ref{Fig:RS}(c)]. In contrast, only the 7.52- and 10.00-THz $A_1$ modes are excited by the pump pulses polarized along the $x$-axis [Fig.~\ref{Fig:RS}(d)], which agrees well with the observation that the line corresponding to the 12.03-THz $A_1(xx)$ phonon is also very weak in the spontaneous RS spectra [Fig.~\ref{Fig:RS}(a)]. Non-polar $B_1$ modes can be excited when $|e_x|^2 \neq |e_y|^2$ (see Table~\ref{Table:driving}). In our experiment this corresponds to the pump pulse polarization making a nonzero angle with the $z$-axis. Indeed, the $B_1$ mode is excited by the pump pulses with azimuthal angles $\theta=0$ and $\ 45^\circ$ [Figs.~\ref{Fig:RS}(d,e)].

Polar $E(y)$ coherent phonon modes are expected to be excited under the condition $e_z e_x^*+e_x e_z^* \neq 0$, which is met at $\theta=45^\circ$. However, as follows from a comparison of the spectra in Figs.~\ref{Fig:RS}(b,e), no line associated with the $E(y)$ phonon modes appear in the FFT spectrum of the ISRS-PD data measured in this geometry. This happens because in the ISRS-PD experiment the detection process is as important as the excitation. In our experimental geometry, optically excited coherent phonons are detected via the modulation of the probe polarization [Eq.~(\ref{eq:oscillations})] originating from transient changes of the linear birefringence. The components of the dielectric tensor modulated by the coherent phonons of a particular symmetry are listed in Table\,\ref{Table:driving}. Clearly, the detection of the $A_1$ and $B_1$ coherent phonons requires the probe polarization to make an angle $\phi\neq0$ and $\ 90^\circ$ (see App.\,\ref{AppI} for the details). In contrast, the modulation of the probe polarization by $E(y)$ coherent phonons vanishes at $\phi=45^\circ$.

The relevant amplitudes $\Delta\eta_0$ [Eq.\,\ref{eq:oscilltions-simple}] of the ellipticity modulation at the phonon frequency $\Omega_0$ for the probe pulses initially polarized at $\phi=45^\circ$ for each pump pulse polarization $\theta$ are
\begin{eqnarray}
\theta&=&0:\label{eq:amplX}\\
A_1&:&\Delta\eta_0=\frac{\pi d_0}{2\lambda_\mathrm{pr}}\mathcal{R}_{xx}(\omega_\mathrm{p})\left[\frac{\mathcal{R}_{zz}(\omega_\mathrm{pr})}{\sqrt{\varepsilon_{zz}}}-\frac{\mathcal{R}_{xx}(\omega_\mathrm{pr})}{\sqrt{\varepsilon_{xx}}}\right];\nonumber\\
B_1&:&\Delta\eta_0=\frac{\pi d_0}{2\lambda_\mathrm{pr}}\frac{\mathcal{R}_{xx}(\omega_\mathrm{p})\mathcal{R}_{xx}(\omega_\mathrm{pr})}{\sqrt{\varepsilon_{xx}}};\nonumber\\
\theta&=&90^\circ:\label{eq:amplZ}\\
A_1&:&\Delta\eta_0=\frac{\pi d_0}{2\lambda_\mathrm{pr}}\mathcal{R}_{zz}(\omega_\mathrm{p})\left[\frac{\mathcal{R}_{zz}(\omega_\mathrm{pr})}{\sqrt{\varepsilon_{zz}}}-\frac{\mathcal{R}_{xx}(\omega_\mathrm{pr})}{\sqrt{\varepsilon_{xx}}}\right];\nonumber\\
\theta&=&45^\circ:\label{eq:amplXZ}\\
A_1&:&\Delta\eta_0=\frac{\sqrt{2}\pi d_0}{4\lambda_\mathrm{pr}}\left(\mathcal{R}_{xx}(\omega_\mathrm{p})+\mathcal{R}_{zz}(\omega_\mathrm{p})\right)\nonumber\\
&&\left[\frac{\mathcal{R}_{zz}(\omega_\mathrm{pr})}{\sqrt{\varepsilon_{zz}}}-\frac{\mathcal{R}_{xx}(\omega_\mathrm{pr})}{\sqrt{\varepsilon_{xx}}}\right];\nonumber\\
B_1&:&\Delta\eta_0=\frac{\sqrt{2}\pi d_0}{4\lambda_\mathrm{pr}}\frac{\mathcal{R}_{xx}(\omega_\mathrm{p})\mathcal{R}_{xx}(\omega_\mathrm{p})}{\sqrt{\varepsilon_{xx}}}.\nonumber\\
E(y)&:&\Delta\eta_0=0.\nonumber
\end{eqnarray}
Note that all $\varepsilon_{ij}$ components should be taken at the optical frequency of the probe pulse here.

It is worth noting that the distinct pump and probe polarizations provide the most efficient conditions for the excitation and detection of coherent phonons or magnons,\cite{Satoh-NPhot2015} as clearly seen from the probe polarization dependence of the ISRS-PD signal shown in Figs.~\ref{Fig:probe_dep}(a,b). We numerically analyzed the dependence of the amplitude of the oscillations of the probe ellipticity measured in ISRS-PD scheme on the probe polarization azimuthal angle, $\phi$, when a coherent phonon of a symmetry $A_1$ modulates the real part of the dielectric permittivity tensor, thus changing the linear crystallographic birefringence. The calculations were performed using the Jones matrix method, taking the experimentally obtained value of 3.1$\times10^{-3}$ of the static birefringence in the $xz$ plane of the studied sample at the probe photon energy. Calculations show [Fig.~\ref{Fig:probe_dep}(c)] that the oscillatory signals are indeed caused by modulation of the linear birefringence of CuB$_2$O$_4$ due to coherent phonons, and demonstrate the importance of the correct choice of the probe polarization for detection of a laser-driven coherent phonon mode of a particular symmetry.

To conclude the discussion of the excitation mechanism of the coherent phonons, we note that ISRS is the sole mechanism of excitation of the $B_1$ coherent phonon mode, while the $A_1$ mode, in general, can be excited via the DECP mechanism as well. The modulation of the probe ellipticity by the coherent phonons excited via ISRS and DECP should possess sine- [Eq.~(\ref{eq:oscillations})] and cosine-like temporal behaviors, respectively.\cite{Ruhman-IEEE1988,Liu-PRL1995} We have therefore fitted the data in Fig.~\ref{Fig:AC_pump} to the sum of two or three damped oscillations.\cite{fit} The resulting initial phases are shown in Table~\ref{Initial_phase_analysis}. All phonon modes show nearly sine-like behaviors, i.e., the initial phases are $\sim 0$ or $\sim 180^\circ$ rather than $\sim 90^\circ$ or $\sim 270^\circ$. We note also that the amplitude of the lines in the FFT spectra show a linear dependence on the pump fluence $I^0_\mathrm{p}$ (see the inset in Fig.~\ref{Fig:RS}), as expected for the coherent phonons excited via ISRS [Eqs.~(\ref{ISRS_ke}--\ref{eq:oscillations})].

\begin{table}
  \caption{Initial phases of the probe ellipticity oscillations $\Delta\eta(t)$ extracted from the fit of the data in Fig.~\ref{Fig:AC_pump} to two or three damped sine-functions.}
  \begin{center}
    \begin{tabular}{c|c|c|c|c}
      \hline \hline
      $\theta$ & 4.38 THz & 7.52 THz & 10.00 THz & 12.03 THz \\ \hline
      $0^\circ$ & (160.6$\pm$4.9)$^\circ$ & (142.9$\pm$4.6)$^\circ$ & (326.7$\pm$0.7)$^\circ$ & - \\
      $45^\circ$ & (186.3$\pm$6.6)$^\circ$ & - & (35.0$\pm$0.6)$^\circ$ & - \\
      $90^\circ$ & - & (358.4$\pm$2.1)$^\circ$ & (13.2$\pm$0.9)$^\circ$ & (199.1$\pm$3.2)$^\circ$ \\
      \hline \hline
      \end{tabular}
  \end{center}
   \label{Initial_phase_analysis}
\end{table}

\subsection{Role of the pump and probe durations in the excitation of coherent phonons via ISRS}\label{Sec-duration}

Now we analyze the significant differences between the amplitudes of the phonon lines in the FFT spectra of the ISRS-PD data and those of the spontaneous RS spectra (Fig.~\ref{Fig:RS}). While both RS and ISRS processes are described by the same Raman tensor components $\mathcal{R}_{ij}$, there is an essential difference between them.\cite{Yan-JCP1987} This is due to the character of the light used in the RS and ISRS experiments; continuous wave versus ultrashort pulse, and due to the different registration techniques. These factors are accounted for in Eqs.~(\ref{ISRS_ke}--\ref{eq:oscillations}) by the constituents (iii--iv) that are dependent on the temporal/spectral profiles of the pump and probe pulses. Furthermore, in the RS experiment light scattering from incoherent phonons takes place in thermal equilibrium, and the Bose-Einstein thermal occupation factor enters the expression for the Raman line intensity.\cite{Loudon-RS1963} This is not the case when the coherent phonons are driven by the ISRS process.\cite{Dougherty-PRB1994}

We consider four $A_1$ and three $B_1$ lowest phonon modes, observed in the spontaneous RS spectra below $15$~THz [Fig.~\ref{Fig:RS}(a,c)].\cite{Pisarev-PRB2013} The lines assigned to these modes are fitted by the sets of Lorentzian functions
\begin{equation}
\mathcal{I}_\mathrm{RS}(\Omega)=c+\sum_{k}\frac{2C_k}{\pi}\frac{\sigma_k}{\frac{1}{\pi^2}(\Omega-\Omega_k)^2+\sigma^2_k},\label{eq:Lorentz}
\end{equation}
where $\Omega_k$ and $\sigma_k$ are the frequency and FWHM of the $k$-th phonon line, respectively. These are listed in Tables~\ref{Table:PhononsA1B1} and \ref{Table:PhononsA1}. The amplitudes $C_k$ of the Stokes lines
 \begin{equation}
C_k\sim\left(N(\Omega_k)+1\right)\left[\mathcal{R}^{\Omega_k}_{ij}(\omega)\right]^2\label{eq:AmplStokes}
\end{equation}
can be used as a measure of the magnitude of the Raman tensor component $\mathcal{R}^{\Omega_k}_{ij}(\omega)$ describing the scattering of light at the frequency $\omega$ by the corresponding phonon. Here $N(\Omega_k)=(\exp(\hbar\Omega_k/k_BT)-1)^{-1}$ is the Bose-Einstein thermal occupation factor. Note that at room temperature the coefficient $N(\Omega_k)+1$ deviates from unity and affects the intensity ratio between different Raman lines when the broad frequency range is considered. As an example, we mention that the ratio $\mathcal{R}^{4.38}/\mathcal{R}^{14.13}$ extracted from the RS data appears to be overestimated by $\sim$25\,\% when the thermal occupation factor is not taken into account.

The peak intensity of the particular phonon line is related to the amplitude and FWHM as $\mathcal{I}^0_\mathrm{RS}(\Omega_k)=2C_k(\pi\sigma_k)^{-1}$. We note that a less intense but broader Raman line may have a larger contribution to the Raman tensor. We also note that the line at $\Omega_k/2\pi=$10.00~THz in the $z(xx)\bar{z}$ spectrum [Fig.~\ref{Fig:RS}(a)] contains contributions from both the $A_1$ and $B_1$ phonons, i.e., $C_{10.00}\sim\left(\mathcal{R}^{10.00}_{xx(A_1)}(\omega)\right)^2+\left(\mathcal{R}^{10.00}_{xx(B_1)}(\omega)\right)^2$. As was found experimentally in Ref.~\onlinecite{Pisarev-PRB2013}, the $A_1$ phonon contribution dominates.

\begin{table}[t]
\caption{Amplitudes $C_k$ and FWHMs $\sigma_k$ of the $A_1+B_1$ phonon lines in the RS ($z(xx)\bar{z}$) and ISRS ($\mathbf{E}^\mathrm{p}\|x$) spectra [Figs.~\ref{Fig:RS}(a,d)]. The Raman tensor $\mathcal{R}_{xx}$ values were extracted from the spontaneous RS data taking into account the thermal occupation numbers $N(\Omega_k)+1$ at $T=293$\,K. The amplitudes of the oscillations $\Delta\eta^\mathrm{calc}_0$ were calculated from the ISRS model, accounting for the pump and probe pulse durations (see text). All amplitudes were normalized to the corresponding amplitude of the 10.00-THz line.}
\begin{tabular}{c|c|c|c|c|c|c}
  \hline\hline
  \multirow{3}{*}{\begin{tabular}{c}$\Omega_k/2\pi$\\(THz)\end{tabular}} & \multicolumn{3}{c|}{RS ($z(xx)\bar{z}$)} & \multicolumn{3}{c}{ISRS ($\mathbf{E}^\mathrm{p}\|x$)}  \\
  \cline{2-7}
   & $C_k$ & $\sigma_k$ & $R_{zz} $ & \begin{tabular}{c}$C_k\sim$\\$\Delta\eta_{0;k}$\end{tabular} & $\sigma_k$ & $\Delta\eta^\mathrm{calc}_{0;k}$ \\
   & (rel.unit) & (GHz) & (rel.unit) & (rel.unit) & (GHz) & (rel.unit) \\ \hline
  4.38 & 0.04 & 68 & 0.16 & 0.1 & 40 & 0.48  \\
  7.52 & 0.02 & 77 & 0.13 & 0.09 & 48 & 0.07  \\
  10.00 & 1 & 77 & 1  & 1 & 79 & 1 \\
  10.71 & 0.08 & 88 & 0.29 & -  & - & 0.04  \\
  12.03 & 0.04 & 108 & 0.21 & - & - & 0.01 \\
  14.13 & 1.31 & 124 & 1.21 & - & - & 0.04  \\
  \hline\hline
\end{tabular}\label{Table:PhononsA1B1}
\end{table}

The RS data reported in Ref.~\onlinecite{Pisarev-PRB2013} were obtained at the photon energy of $\hbar\omega=2.41$~eV. However, as demonstrated in Ref.~\onlinecite{Ivanov-PRB2013}, the change of the excitation energy in the RS experiments from 2.71 to 1.96~eV did not yield any significant redistribution of the intensity of the phonon lines in the spontaneous RS spectra. This observation indicates that the RS process at the photon energies as high as 2.71~eV is of a non-resonant nature. Therefore, we can consider the relative intensities of the Raman lines reported in Ref.~\onlinecite{Pisarev-PRB2013} as a reliable measure of the Raman tensor components for the case of non-resonant scattering at the pump ($\hbar\omega_\mathrm{p}=1.08$~eV) and probe ($\hbar\omega_\mathrm{p}=1.55$~eV) photon energies used in our experiments: $\mathcal{R}^{\Omega_{k}}_{ij}(\omega_\mathrm{p})=\mathcal{R}^{\Omega_{k}}_{ij}(\omega_\mathrm{pr})$.

\begin{table}[t]
\caption{Amplitudes $C_k$ and FWHMs $\sigma_k$ of the $A_1$ phonon lines in the RS ($x(zz)\bar{x}$) and ISRS ($\mathbf{E}^\mathrm{p}\|z$) spectra [Figs.~\ref{Fig:RS}(c,f)]. The Raman tensor $\mathcal{R}_{zz}$ values were extracted from the spontaneous RS data taking into account the thermal occupation numbers $N(\Omega_k)+1$ at $T=293$\,K. The amplitudes of the oscillations $\Delta\eta^\mathrm{calc}_0$ were calculated from the ISRS model, accounting for the pump and probe pulse durations (see text). All amplitudes were normalized to the corresponding amplitude of the 10.00-THz line.}
\begin{tabular}{c|c|c|c|c|c|c}
  \hline\hline
  \multirow{3}{*}{\begin{tabular}{c}$\Omega_k/2\pi$\\(THz)\end{tabular}} & \multicolumn{3}{c|}{RS ($x(zz)\bar{x}$)} & \multicolumn{3}{c}{ISRS ($\mathbf{E}^\mathrm{p}\|z$)}  \\
  \cline{2-7}
   & $C_k$ & $\sigma_k$ & $R_{zz}$ & \begin{tabular}{c}$C_k\sim$\\$\Delta\eta_{0;k}$\end{tabular} & $\sigma_k$ & $\Delta\eta^\mathrm{calc}_{0;k}$ \\
   & (rel.unit) & (GHz) & (rel.unit) & (rel.unit) & (GHz) & (rel.unit) \\ \hline
  7.52 & 0.09 & 54 & 0.25 & 0.39 & 56 & 0.35 \\
  10.00 & 1 & 57 & 1 & 1 & 70 & 1 \\
  12.03 & 1.2 & 75 & 0.88 & 0.24 & 77 & 0.23 \\
  14.13 & 3.12 & 115 & 1.65 & - & - & 0.08  \\
  \hline\hline
\end{tabular}\label{Table:PhononsA1}
\end{table}

The normalized amplitudes of the phonon lines in the FFT spectra obtained for the pump-probe traces [Figs.~\ref{Fig:RS}(d--f)] using the fit function [Eq.~(\ref{eq:Lorentz})] are summarized in Tables~\ref{Table:PhononsA1B1} and \ref{Table:PhononsA1}. The relation between the Lorentzian line amplitudes $C_i\sim\Delta\eta_0$ and the Raman tensor components can be obtained from Eqs.~(\ref{eq:amplX}--\ref{eq:amplXZ}) and is, in general, less straightforward than in the case of RS because of the chosen detection technique. For the sake of simplicity, we assume the ratio $\mathcal{R}_{zz}/\mathcal{R}_{xx}\approx3$ for the $A_1$ mode of frequency $\Omega/2\pi=$7.52~THz, as estimated from Eqs.~(\ref{eq:amplX}),(\ref{eq:amplZ}) and from the data in Figs.~\ref{Fig:RS}(d,f).
Then Eqs.~(\ref{eq:amplX}),(\ref{eq:amplZ}) are reduced to
\begin{eqnarray}
\theta=0&:&\label{eq:amplX-simple}\\
A_1&:&C_i\sim\Delta\eta_0\sim\mathcal{R}^2_{xx}\left[\frac{3}{\sqrt{\varepsilon_{zz}}}-\frac{1}{\sqrt{\varepsilon_{xx}}}\right];\nonumber\\
B_1&:&C_i\sim\Delta\eta_0\sim\frac{\mathcal{R}^2_{xx}}{\sqrt{\varepsilon_{xx}}}.\nonumber\\
\theta=90^\circ&:&\label{eq:amplZ-simple}\\
A_1&:&C_i\sim\Delta\eta_0\sim\mathcal{R}^2_{zz}\left[\frac{1}{\sqrt{\varepsilon_{zz}}}-\frac{1}{3\sqrt{\varepsilon_{xx}}}\right].\nonumber
\end{eqnarray}

\begin{figure}[t]
\includegraphics[width=8.6cm]{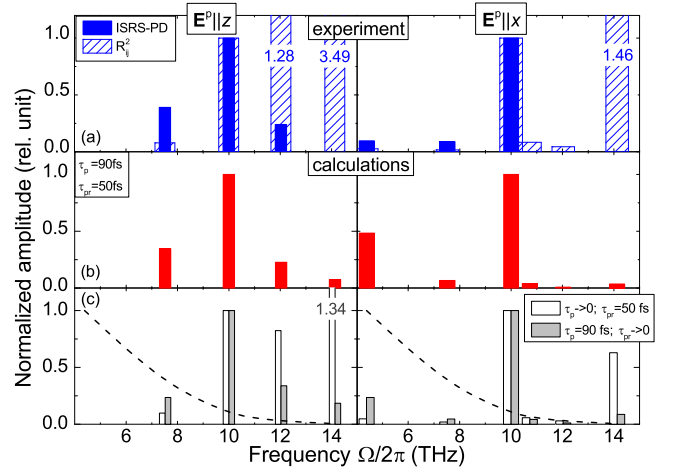}
\caption{(Color online) (a) Amplitudes $C_k\sim\Delta\eta_0$ (blue bars) of the coherent phonon modes, as extracted from a FFT of the ISRS-PD experimental data [Figs.~\ref{Fig:RS}(d--f)]. Also shown (hatched bars) are the values of the corresponding squared Raman tensor components $\mathcal{R}^2_{ij}$ (see Tables~\ref{Table:PhononsA1B1} and \ref{Table:PhononsA1}). (b--c) Amplitudes of the probe ellipticity $\Delta\eta_0$ calculated using Eq.~(\ref{eq:oscillations}) for the cases (b) $\tau_\mathrm{pr}$=50~fs, $\tau_\mathrm{p}=$90~fs, (c) $\tau_\mathrm{pr}$=50~fs, $\tau_\mathrm{p}\rightarrow$0 (open bars), and $\tau_\mathrm{pr}\rightarrow$0, $\tau_\mathrm{p}$=90~fs (gray bars). All amplitudes are normalized by the amplitude of the 10.00-THz mode in the relevant spectra. Dashed lines in the panel (c) show the frequency dependence of the normalized coefficient $\alpha(90~\mathrm{fs},\Omega)$.}
\label{Fig:RSvsISRS}
\end{figure}

From Eqs.~(\ref{eq:amplX-simple}),(\ref{eq:amplZ-simple}) it follows that the amplitudes $C_i$ obtained for the ISRS-PD data should be compared with those from the RS spectra. In Fig.~\ref{Fig:RSvsISRS}(a) the amplitudes $C_i$ that are plotted are normalized with respect to the 10.00-THz line in the relevant spectra. There are two general trends. First, the 10.00- and 12.03-THz $A_1$ phonon lines in the $x(zz)\bar{x}$ RS spectrum have similar amplitudes, while in the ISRS spectrum the amplitude of the 12.03-THz coherent phonon line excited by the pump pulses with $\mathbf{E}^\mathrm{p}\|z$ appears to be $\sim$4 times smaller than that of the 10.00-THz line. Furthermore, no coherent phonons with frequencies above 12.03-THz could be detected in the ISRS-PD experiment, while some of the phonon lines in this range gain high intensities in the RS spectra, e.g., the 14.13-THz line. Second, from the RS data it follows that $\left(\mathcal{R}^{10.00}_{zz}/\mathcal{R}^{7.52}_{zz}\right)^2\sim10$ for the $A_1$ modes. In the ISRS spectrum, the corresponding ratio appears to be only $\Delta\eta_0^{10.00}/\Delta\eta_0^{7.52}\sim3$.

To analyze the observed difference in the RS and ISRS spectra we consider how the durations of both the pump and probe pulses affect the results. We assume that the pump and probe pulses have Gaussian temporal profiles;
\begin{equation}
\mathcal{E}(t)=\mathcal{E}_0e^{-2\ln2\left(t^2/\tau^2\right)}.\label{eq:profileT}
\end{equation}
In the frequency domain, both these pulses are assumed to be Fourier-limited, and their spectral profiles are obtained as a Fourier transform of Eq.\,(\ref{eq:profileT})
\begin{equation}
\mathcal{E}(\omega_\mathrm{p(pr)})=\mathcal{E}_0\frac{\sqrt{2\ln2}}{\sqrt{\pi}\sigma_\mathrm{p(pr)}}e^{-2\ln2\left(\omega^2/\sigma_\mathrm{p(pr)}^2\right)},\label{eq:profileF}
\end{equation}
where $\sigma_\mathrm{p(pr)}=4\ln2\,\tau^{-1}_\mathrm{p(pr)}$ is the FWHM of the pump and probe pulse intensity profiles in the spectral domain.

The amplitude $\Delta\eta_0$ of the measured probe polarization oscillations in the ISRS-PD experiment increased as the probe pulse duration $\tau_\mathrm{pr}$ decreased, as compared with the period of a particular coherent phonon. To illustrate this effect we calculated the expected change $\Delta\eta(t)$ in the ISRS-PD experiment using Eqs.~(\ref{eq:oscillations}),(\ref{eq:amplX-simple}), and (\ref{eq:amplZ-simple}), as well as assuming infinitesimally short pump pulses, i.e., $\alpha(\tau_\mathrm{p}\rightarrow0)=1$. In Fig.~\ref{Fig:RSvsISRS}(c) we plot the normalized FFT intensities of $A_1$ ($\mathbf{E}^\mathrm{p}\|z$), $A_1$ and $B_1$ ($\mathbf{E}^\mathrm{p}\|x$) coherent phonons calculated in this way. The probe duration was taken to be $\tau_\mathrm{pr}$=50~fs ($\sigma_\mathrm{pr}/2\pi$=9~THz), i.e., equal to that used in our experiments. The resulting 12.03-THz line ($\tau_\mathrm{pr}\Omega_i\approx0.7$) is noticeably suppressed as compared with the 10.00-THz line ($\tau_\mathrm{pr}\Omega_i\approx0.5$). This is despite the fact that the intensities of these two phonon modes in the $x(zz)\bar{x}$ RS spectra (Fig.~\ref{Fig:RSvsISRS}(a)) and, consequently, the corresponding Raman tensor components, are nearly equal. The effect of the probe duration is even more pronounced when the ratios between the intensities of the 10.00- and 14.13-THz lines in the RS spectrum are compared with those in the calculated ISRS spectrum. Thus, considering only the probe pulse duration, one can account for, to a certain extent, the differences in the intensities of the phonon lines in the RS and ISRS spectra in the high-frequency part of the spectrum.

While the duration of the probe pulse affects the detection part of the pump-probe experiment, the duration of the pump pulse determines the efficiency of the coherent phonon excitation. This is accounted for by the coefficient $\alpha_k(\tau_\mathrm{p},\Omega_0)$ (\ref{eq:alpha}), which has the following form for the Gaussian pulses:
\begin{equation}
\alpha_k(\tau_\mathrm{p},\Omega_0)=e^{-2\ln2\left(\Omega_0^2/\sigma_\mathrm{p}^2\right)}.
\label{eq:alpha-Gauss}
\end{equation}
The normalized coefficient $\alpha(90~\mathrm{fs},\Omega)$ in the range below 15~THz is shown in Fig.~\ref{Fig:RSvsISRS}(c) using the dashed lines. This graph clearly demonstrates that the efficiency of the excitation of, e.g., the 10.00-THz phonon mode would be $\sim$3.5 times greater than that of the 12.03-THz mode, given equal Raman tensor components associated with these modes. We illustrate the effect of the pump pulse duration on the outcome of the pump-probe experiments by calculating the ISRS-PD spectra using Eqs.~(\ref{eq:oscillations}),(\ref{eq:amplX-simple}),(\ref{eq:amplZ-simple}), and(\ref{eq:alpha-Gauss}), as well as assuming an infinitesimally short probe pulse $I_\mathrm{pr}(t)=I^0_\mathrm{pr}\delta(t)$ and a pump pulse of duration $\tau_\mathrm{p}=$90~fs ($\sigma_\mathrm{p}/2\pi\approx$5~THz). We note that the amplitudes $\Delta\eta_0$ obtained at $\tau_\mathrm{pr}\to0$ are proportional to the squared amplitudes of the normal coordinates $Q_0$, \textit{i.e.}, to the corresponding atomic displacements excited by the pump pulse.

Finally, we calculate the ISRS-PD amplitudes of the phonons by considering the durations of both the pump ($\tau_\mathrm{p}$=90~fs) and probe ($\tau_\mathrm{pr}$=50~fs) pulses. The results are shown in Fig.~\ref{Fig:RSvsISRS}(b) and listed in Tables~\ref{Table:PhononsA1B1} and \ref{Table:PhononsA1}. For the excitation of the $A_1$ mode (left side of Fig.~\ref{Fig:RSvsISRS}(b)), our model adequately describes the main trends observed in the experiments. As shown, for the three $A_1$ modes with frequencies lying in the range 7--15~THz, this model accounts for either the partial or total suppression of the phonon lines with frequencies above 10~THz in the ISRS-PD spectrum, in all considered geometries. In the lower frequency range the discrepancy between the experimental results and model occurs only for the 4.38-THz $B_1$ phonon mode, in which the amplitude appears to be overestimated. Importantly, our calculations show that a reasonable agreement between the calculated ISRS-PD spectra and those obtained from the pump-probe experiment is obtained only by including the durations of both the pump and probe pulses.

To complete the comparative analysis of the spontaneous RS and ISRS-PD data, we note that the phonon lines in the RS spectra have very narrow widths (see Tables~\ref{Table:PhononsA1B1} and \ref{Table:PhononsA1}).\cite{Pisarev-PRB2013} Thus, in the $x(zz)\bar{x}$ RS spectra, the line at 10.00~THz is characterized by a FWHM of $\approx$60~GHz. In the FFT spectra of the ISRS-PD data measured with $\mathbf{E}^\mathrm{p}\|z$, this line has a FWHM of 70~GHz, i.e., broader by only $\sim15\%$. We note that, as discussed in Ref.~\onlinecite{Misochko-JETP2001}, the shorter decay time (broader line width) of the coherent phonons excited via ISRS, as compared to that of incoherent phonons contributing to the RS spectra, is caused by different energy dissipation channels. However, some of the lines appear to be even narrower in the ISRS-PD experiments compared with those of the RS spectra. This is most probably related to the experimental limitations in the temporal and spectral resolutions in these two cases.

\section{Conclusions}\label{Sec-conclusion}

We have performed a detailed study of laser-induced excitation and detection of multiple coherent phonon modes in the ISRS-PD experiment in the dielectric copper metaborate CuB$_2$O$_4$, characterized by a large primitive unit cell containing 42 atoms. In total, three non-polar $A_1$ and one non-polar $B_1$ modes were distinguished in the frequency range of 4--13~THz. We have shown that $90$-fs linearly polarized laser pulses with a central photon energy in the optical transparency range ($\hbar\omega_\mathrm{p}$=1.08~eV) excite the coherent phonons via ISRS. By comparing the results of the ISRS-PD experiment to spontaneous RS spectra in this material, we demonstrated that the relationship between the amplitudes and intensities of the phonon lines in these two types of experiments, is determined by both the excitation and detection conditions, respectively. Namely, the amplitude of the excited coherent phonon is determined by the polarization, intensity, and duration of the pump pulses. Probe pulse polarization and duration are as important as those of the pump pulses. They determine how the transient changes of the dielectric permittivity due to excited coherent phonons lead to the modulation of the probe polarization detected in the ISRS-PD experiment.

A comparison between the spontaneous RS spectra and ISRS-PD data also allowed us to analyze in detail the limitations imposed by the durations of the pump and probe pulses on the excitation and detection of the coherent phonons. Accounting for both durations is required to adequately calculate the modulation of the probe polarization in the pump-probe experiment using the spontaneous RS data. It is important to understand the role played by the pump and probe pulse durations in ISRS, because ever-shorter laser pulses are currently employed in pump-probe experiments, providing access to high-energy collective excitations in solids. To the best of our knowledge, no such detailed analysis for the case of multiple coherent phonon modes excitation has been reported to date.

Finally, we would like to note that the reported details of the ultrafast coherent lattice dynamics in copper metaborate CuB$_2$O$_4$ are of importance in light of recent attention given to the high-frequency phonon-magnon interaction and their role in the ultrafast dynamics driven by femtosecond laser pulses.\cite{Korenev-Nphys2016,Nova-NPhys2016} We demonstrated the excitation of coherent vibrations of Cu$^{2+}$ ions belonging to different magnetic sublattices. In particular, the $B_1$ 4.38-THz coherent phonon is the vibration in a Cu($4b$)O$_4$ complexes solely with periodic displacement of Cu$^{2+}(4b)$ ions along the $z$-axis. Magnetic ions in this complex provide the strongest contribution to the magnetic ordering in CuB$_2$O$_4$.

\section{Acknowledgements}

We thank V. Yu.\ Davydov for help with the spontaneous Raman scattering experiments. AMK acknowledges support from the Japanese Society for Promotion of Sciences (JSPS) via the Short-Term Fellowship Program for European and North-American young researchers during her stay at the University of Tokyo. TS was supported by JSPS KAKENHI (No.\ JP15H05454 and JP26103004) and JSPS Core-to-Core Program (A. Advanced Research Networks). RVP acknowledges the support from the Russian Science Foundation (grant No.\ 16-12-10456).

\appendix
\section{Probe polarization changes in the ISRS-PD experiment}\label{AppI}

Here we derive the expression that relates the changes of the ellipticity of the probe pulses measured in the ISRS-PD experiment to the Raman tensor components, and the pump and probe parameters.

\subsection{A case of $A_1$ and $B_1$ coherent phonons}
Let the coherent phonon only contribute to the modulation of the diagonal components of the dielectric permittivity tensor, as in the case of the $A_1$ or $B_1$ phonons. Then, the expression for the dielectric tensor is $\varepsilon_{ij}+\delta\varepsilon_{ij}(t)$, where $\delta\varepsilon_{ij}(t)=0$ if $i\neq j$. $\delta\varepsilon_{ij}$ is related to the coherent phonon normal coordinate as $\delta\varepsilon_{ij}(t)=\mathcal{R}_{ij}Q(t)$.

We consider the light propagating along the $j$-axis. The eigenwaves in this case are two orthogonal linearly-polarized waves $E_{i}$ and $E_{k}$, where $\mathbf{i}$ and $\mathbf{k}$ are the unit vectors in the directions of the $i$- and $k$-axes, respectively. The corresponding complex refraction indices for these eigenwaves are $n_{i(k)}-\mathrm{i}\kappa_{i(k)}=\sqrt{\varepsilon_{ii(kk)}+\delta\varepsilon_{ii(kk)}}$. After travelling the distance $d_0$ these waves acquire additional phases of $2\pi n_{i(k)}d_0/\lambda$ and their amplitudes are decreased by $e^{-2\pi\kappa_{i(k)}d_0/\lambda}$.

First, for the sake of clarity we consider a nondissipative medium \textbf{($\kappa_{i(k)}=0$)}. Therefore, the diagonal components of the dielectric permittivity tensor and, consequently, refractive indices, are real values. We assume that the light is initially linearly polarized at an angle $\phi$ to the $i$-axis [Fig.\,\ref{Fig:experimental}(a)]. Then, upon traveling the distance $d_0$, the light becomes elliptically polarized. Considering that the relationship between the complex amplitudes of the orthogonal components of the light polarization is given by\cite{Azzam}
\begin{equation}
E_k/E_i=\frac{\tan\phi+\mathrm{i}\tan\eta}{1-\mathrm{i}\tan\phi\tan\eta},\nonumber
\end{equation}
we obtain the ellipticity $\eta(t)$ [Fig.\,\ref{Fig:experimental}(b)] that relates the acquired phase shift between two eigenwaves by \begin{equation}
\tan2\eta(t)=\frac{2\pi(n_i(t)-n_k(t))d_0}{\lambda}\sin{2\phi}.\label{AppIeq:eliipticity-general}
\end{equation}
For small $\eta(t)$ this yields
\begin{eqnarray}
\eta+\Delta\eta(t)&=&\frac{\pi d_0}{\lambda}(n_i(t)-n_k(t))\sin{2\phi}\approx\label{AppIeq:ellipticity}\\
&&\frac{\pi d_0}{\lambda}\left(\sqrt{\varepsilon_{ii}}-\sqrt{\varepsilon_{kk}}\right)\sin{2\phi}+\nonumber\\
&&\frac{\pi d_0}{2\lambda}\left(\frac{\delta\varepsilon_{ii}(t)}{\sqrt{\varepsilon_{ii}}}-\frac{\delta\varepsilon_{kk}(t)}{\sqrt{\varepsilon_{kk}}}\right)\sin{2\phi},\nonumber
\end{eqnarray}
where $\eta$ is the static contribution to the ellipticity due to crystallographic birefringence. This expression was obtained by taking into account that the modulation of the dielectric permittivity tensor induced by coherent phonons is significantly weaker than the value of the corresponding unperturbed component, \textit{i.e.} $\delta\varepsilon_{ij}\ll\varepsilon_{ij}$.

As can be seen from Eq.~(\ref{AppIeq:ellipticity}), there are two contributions to the ellipticity of light passing though the medium. The first one is related to the birefringence of the unperturbed medium and does not contribute to the measured ISRS-PD signal. Thus, the changes in the ellipticity are related to the normal coordinate of the corresponding coherent phonon by
\begin{eqnarray}
\Delta\eta(t)\approx\frac{\pi d_0}{2\lambda}\left(\frac{\mathcal{R}_{ii}}{\sqrt{\varepsilon_{ii}}}-\frac{\mathcal{R}_{kk}}{\sqrt{\varepsilon_{kk}}}\right)Q(t)\sin{2\phi},\label{AppIeq:ellipticity-fin-angle}
\end{eqnarray}
where we took into account the relationship between the dielectric permittivity tensor components and phonon normal coordinate.

As can be seen from Eq.\,(\ref{AppIeq:ellipticity-fin-angle}), in order to detect the $A_1$ and $B_1$ coherent phonons by measuring the changes of the probe pulses ellipticity $\Delta\eta(t)$ the geometry must be chosen with $\phi\neq0,\,90^\mathrm{o}$, and $\phi=45^\mathrm{o}$ ensuring the best sensitivity. It also follows from Eq.\,(\ref{AppIeq:ellipticity-fin-angle}), and the Table\,\ref{Table:driving}, that the $A_1$ coherent phonons would not manifest themselves in the experiments with pump or probe light propagating along the $z$-axis of the crystal, since $\delta\varepsilon_{xx}=\delta\varepsilon_{yy}$.

In our experiments, the QWP was placed after the sample with its axis  parallel to the incoming probe beam polarization. It is convenient to consider this experimental geometry in the coordinate frame, with two axes directed along the light propagation direction and incoming light polarization. In this frame the Jones vector for the light passing through the medium, with an acquired ellipticity $\Delta\eta(t)$, has the form $[E_0; E_0\Delta\eta(t)e^{i\mathrm \pi/2}]$. After passing through the QWP, which introduces a $\pi/2$ phase shift between the components of the Jones vector, the light becomes linearly polarized with the Jones vector $[E_0; E_0\Delta\eta(t)]$. Thus, the azimuthal angle of the probe pulses appear to be rotated by the angle $\Delta\phi(t)\approx\tan(\Delta\phi(t))=\Delta\eta(t)$.

\subsection{A case of $B_2$, $E(x)$, and $E(y)$ coherent phonons}

For the case of a coherent phonon mode, which modulates the off-diagonal components of the dielectric permittivity tensor $\delta\varepsilon_{ik}$, e.g., $E(y)$, the eigenwaves are two linearly polarized waves with azimuthal angles $\arctan{(\delta\varepsilon_{ik}/(\varepsilon_{kk}-\varepsilon_{ii}))}$ with the $i$- and $k$-axes. Here we assumed that the changes of the dielectric permittivity tensor components are smaller than the difference between the diagonal tensor components of the unperturbed medium. The corresponding refractive indices are $n_{1(2)}=\sqrt{\varepsilon_{ii(kk)}\pm(\delta\varepsilon_{ik}(t))^2/(\varepsilon_{kk}-\varepsilon_{ii})}$ ($\kappa_{i(k)}=0$ for a nondissipative medium).

The detailed expression for the changes of the light polarization are rather complex in this case, and we simplify it by noting the following. In contrast to the previously considered case, here the excited coherent phonons perturb the basis formed by the eigenwaves, while the refractive indices for the eigenwaves can be treated as unchanged $n_{1(2)}\approx\sqrt{\varepsilon_{ii(kk)}}$. Therefore, it is convenient to consider this scenario as if the basis remains unchanged and coincides with the $i$- and $k$-axes, but the light polarization azimuthal angle is modulated by the angle $\arctan{(\delta\varepsilon_{ik}(t)/(\varepsilon_{kk}-\varepsilon_{ii}))}\approx\delta\varepsilon_{ik}(t)/(\varepsilon_{kk}-\varepsilon_{ii})$. By replacing $\phi$ by $\phi+\delta\varepsilon_{ik}(t)/(\varepsilon_{kk}-\varepsilon_{ii})$ in Eq.\,(\ref{AppIeq:eliipticity-general}) we obtain the expression for the oscillatory part of the probe ellipticity:
\begin{eqnarray}
\Delta\eta(t)&\approx&\frac{\pi d_0}{\lambda}\frac{2\mathcal{R}_{ik}}{\sqrt{\varepsilon_{kk}}+\sqrt{\varepsilon_{ii}}}Q(t)\cos{2\phi}+\label{AppIeq:ellipticity-Ey}\\
&&\frac{2\pi d_0}{\lambda}\left(\frac{\mathcal{R}_{ik}Q(t)}{\varepsilon_{kk}-\varepsilon_{ii}}\right)^2(\sqrt{\varepsilon_{kk}}+\sqrt{\varepsilon_{ii}})\sin2\phi.\nonumber
\end{eqnarray}
Thus, in contrast to the $A_1$ or $B_1$ coherent phonons, detection of the, e.g., $E(y)$, coherent phonons can be realized with probe pulses polarized at $\phi=0$, while at $\phi=45^\mathrm{o}$ their effect on the probe polarization is quadratic on small perturbations.

\begin{figure}[t]
\includegraphics[width=8.6cm]{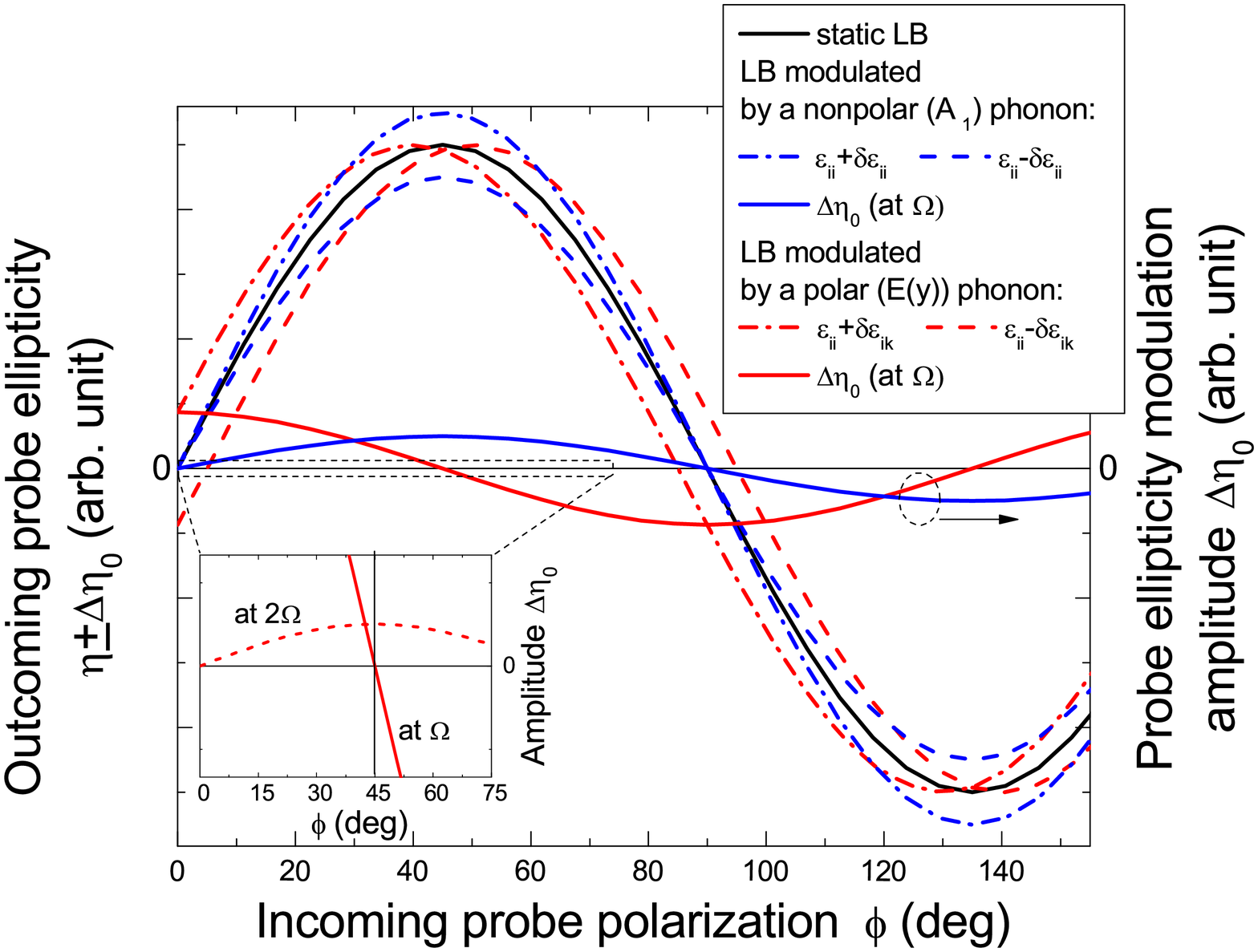}
\caption{(Color online) Ellipticity of the probe polarization $\eta\pm\delta\eta_0$ occurring due to static linear birefringence (LB) (black solid line), and modulation of the dielectric tensor by non-polar (red lines) and polar (blue lines) coherent phonons. The dashed and dash-dotted lines capture the ellipticity at maximal positive and negative changes of the dielectric permittivity by coherent phonons, respectively. The red and blue solid lines show the expected amplitudes of the ellipticity modulation by non-polar and polar coherent phonons at the frequency $\Omega$. The inset displays a magnification of the range close to $\phi=45^\mathrm{o}$ to show that only a weak modulation of the ellipticity at the frequency $2\Omega$ is expected for this polarization for the case of a polar coherent phonon.}
\label{Fig:AppPolDep}
\end{figure}

To summarize, we plot in Fig.\,\ref{Fig:AppPolDep} the ellipticity of the probe pulses as a function of incoming polarization $\phi$ for two considered cases. If the non-polar phonons modulate the diagonal elements of the dielectric permittivity tensor, this modulation manifests itself in a periodic change of the amplitude of the $\eta$ versus $\phi$ dependence, while the knots and maxima of the dependence remain at their positions. The amplitude of the modulation $\Delta\eta_0$ increases as the incoming polarization reaches $45^\mathrm{o}$, and is zero if $\phi=0,\,90^\mathrm{o}$. In contrast, the polar coherent phonon periodically shifts the knots and maxima of the $\eta$ versus $\phi$  dependence, leaving its amplitude unchanged. As a result, the amplitude of the ellipticity modulation $\Delta\eta_0$ at the phonon frequency $\Omega$ reaches its maximum at $\phi=0,\,90^\mathrm{o}$. At $\phi=45^\mathrm{o}$ only a weak modulation at doubled phonon frequency $2\Omega$ is expected (see inset in Fig.\,\ref{Fig:AppPolDep}).

\subsection{A case of opaque medium}

When the absorption in a medium cannot be neglected, the diagonal dielectric tensor components are complex ($\kappa_{i(k)}>0$). We consider here the limiting case when the imaginary part of the dielectric permittivity dominates, and the $A_1$ coherent phonon of a symmetry is detected. Then the light propagating a distance $d_0$ acquires the modulated change of the azimuthal angle of:
\begin{equation}
\Delta\phi(t)\approx\frac{\pi d_0}{2\lambda}\left(\frac{\mathcal{R}_{ii}}{\sqrt{\varepsilon_{ii}}}-\frac{\mathcal{R}_{kk}}{\sqrt{\varepsilon_{kk}}}\right)Q(t)\sin{2\phi},\label{AppIeq:rotation}
\end{equation}
Thus, the coherent phonon mode can be detected by directly measuring the rotation of the probe polarization passing through the crystal. We note here that the rotation (Eq.~(\ref{AppIeq:rotation})) induced by coherent phonons via linear dichroism is sensitive to the incoming probe polarization in the same way as the ellipticity occurring due to linear birefringence.


\begin{thebibliography}{99}

\bibitem{deSilvestri-CPL1985} S. De Silvestri, J. G. Fujimoto, E.P. Ippen, E. B. Gamble Jr., L. R. Williams, and K. A. Nelson, 
    Chem.\ Phys.\ Lett.\ \textbf{116}, 146 (1985).
\bibitem{Dhar-ChemRev1994} L. Dhar, J. A. Rogers, and K. A. Nelson, Chem.\ Rev.\ \textbf{94}, 157 (1994).
\bibitem{Merlin-SSC1997} R. Merlin, Solid State Commun.\ \textbf{102}, 207 (1997).
\bibitem{Yan-JCP1987} Y.-X. Yan and K. A. Nelson, 
J. Chem. Phys. \textbf{87}, 6257 (1987).
\bibitem{Misochko-JETP2001} O. V. Misochko, 
Zh.\ Eksp.\ Teor.\ Fiz.\ \textbf{119}, 285 (2001) [J. Exp.\ Theor.\ Phys.\ \textbf{92}, 246 (2001)].
\bibitem{Kirilyuk-RMP2010} A. Kirilyuk, A. V. Kimel, and Th.\ Rasing, 
Rev.\ Mod.\ Phys.\ \textbf{82}, 2731 (2010).
\bibitem{Kalashnikova-PhysUsp2015} A. M. Kalashnikova, A. V. Kimel, and R. V. Pisarev, 
Usp.\ Fiz.\ Nauk \textbf{185}, 1064 (2015) [Phys.-Usp.\ \textbf{58}, 969 (2015)].
\bibitem{Bakker-RMP1998} H. J. Bakker, S. Hunsche, and H. Kurz, 
Rev.\ Mod.\ Phys.\ \textbf{70}, 523 (1998).
\bibitem{Lorenzana-EPJ2013} J. Lorenzana, B. Mansart, A. Mann, A. Odeh, M. Chergui, and F. Carbone, 
Eur.\ Phys.\ J. Special Topics \textbf{222}, 1223 (2013).
\bibitem{Weiner-Science1990} A. M. Weiner, D. E. Leaird, G. P. Wiederrecht, and K. A. Nelson, 
    Science \textbf{247}, 1317 (1990).
\bibitem{Weiner-JOSAB1991} A. M. Weiner, D. E. Leaird, G. P. Wiederrecht, and K. A. Nelson, 
    J. Opt.\ Soc.\ Amer.\ B \textbf{8}, 1264 (1991).
\bibitem{Kawashima-ARPC1995} H. Kawashima, M. M. Wefers, and K. A. Nelson, 
Annu.\ Rev.\ Phys.\ Chem.\ \textbf{46}, 627 (1995).
\bibitem{Feurer-Science2003} T. Feuer, J. C. Vaughan, and K. A. Nelson, 
Science \textbf{299}, 374 (2003).
\bibitem{McGrane-NJP2009} S. D. McGrane, R. J. Scharff, M. Greenfield, and D. S. Moore, 
    New J.\ Phys.\ \textbf{11}, 105047 (2009).
\bibitem{Shimada-APL2012} T. Shimada, Ch.\ Frischkorn, M. Wolf, ana T. Kampfrath, 
J.\ Appl.\ Phys.\ \textbf{112}, 113103 (2012).
\bibitem{Zeiger-PRB1992} H. J. Zeiger, J. Vidal, T. K. Cheng, E. P. Ippen, G. Dresselhaus, and M. S. Dresselhaus, Phys.\ Rev.\ B \textbf{45}, 768 (1992).
\bibitem{Bossini-PRB2014} D. Bossini, A. M. Kalashnikova, R. V. Pisarev, Th.\ Rasing, and A. V. Kimel, 
    Phys.\ Rev.\ B \textbf{89}, 060405(R) (2014).
\bibitem{Mann-PRB2015} A. Mann, E. Baldini, A. Tramontana, E. Pomjakushina, K. Conder, Ch.\ Arrell, F. van Mourik, J. Lorenzana, and F. Carbone, 
    Phys.\ Rev.\ B \textbf{92}, 035147 (2015).
\bibitem{Ishioka-JPCM2013} K. Ishioka, K. Kato, N. Ohashi, H. Haneda, M. Kitajima, and H. Petek, 
    J. Phys.\ Condens.\ Matter \textbf{25}, 205404 (2013).
\bibitem{Li-PRL2013} J. J. Li, J. Chen, D. A. Reis, S. Fahy, and R. Merlin, 
    Phys.\ Rev.\ Lett.\ \textbf{110}, 047401 (2013).
\bibitem{Nakamura-PRB2015} K. G. Nakamura, Y. Shikano, and Y. Kayanuma, 
    Phys.\ Rev.\ B \textbf{92}, 144304 (2015).
\bibitem{Misochko-JETP2016} O. V. Misochko, 
    Zh.\ Eksp.\ Teor.\ Fiz.\ \textbf{150}, 37 (2016) [J. Exp.\ Theor.\ Phys.\ \textbf{123}, 292 (2016)].
\bibitem{Bossini-NComm2015} D. Bossini, S. Dal Conte, Y. Hashimoto, A. Secchi,	R. V. Pisarev, Th.\ Rasing, G. Cerullo, and A. V. Kimel, 
    Nature Commun.\ \textbf{7}, 10645 (2015).
\bibitem{Brida-JOpt2010} D. Brida, C. Manzoni, G. Cirmi, M. Marangoni, S. Bonora, P. Villoresi, S. De Silvestri, and G. Cerullo, 
    J. Opt.\ \textbf{12}, 013001 (2010).
\bibitem{Satoh-NPhot2015} T. Satoh, R. Iida, T. Higuchi, M. Fiebig, and T.\ Shimura, 
    Nature Photon.\ \textbf{9}, 25 (2015).
\bibitem{Bardeen-PRL1995} C. J. Bardeen, Q. Wang, and C. V. Shank, Phys.\ Rev.\ Lett.\ \textbf{75}, 3410 (1995).
\bibitem{Wand-PCCP2010} A. Wand, Sh. Kallush, O. Shoshanim, O. Bismuth, R. Kosloff, and S. Ruhman, Phys. Chem. Chem. Phys. \textbf{12}, 2149 (2010).
\bibitem{Martinez-R-Acta1971} M. Martinez-Ripoll, S. Martinez-Carrera, and S. Garcia-Blanco, Acta\ Crystallogr. B \textbf{27}, 677 (1971).
\bibitem{Pisarev-PRB2010} R. V. Pisarev, A. M. Kalashnikova, O. Sch\"{o}ps, and L. N. Bezmaternykh, Phys.\ Rev.\ B \textbf{84}, 075160 (2011).
\bibitem{Roessli-PRL2001} B. Roessli, J. Schefer, G. A. Petrakovskii, B. Ouladdiaf, M. Boehm, U. Staub, A. Vorotinov, and L. Bezmaternikh, Phys.\ Rev.\ Lett.\ \textbf{86}, 1885 (2001).
\bibitem{Saito-PRL2008} M. Saito, K. Ishikawa, K. Taniguchi, and T. Arima, Phys.\ Rev.\ Lett.\ \textbf{101}, 117402 (2008).
\bibitem{Boldyrev-PRL2015} K. N. Boldyrev, R. V. Pisarev, L. N. Bezmaternykh, and M. N. Popova, Phys.\ Rev.\ Lett.\ \textbf{114}, 247210 (2015).
\bibitem{Pisarev-PRB2013} R. V. Pisarev, K. N. Boldyrev, M. N. Popova, A. N. Smirnov, V. Yu.\ Davydov, L. N. Bezmaternykh, M. B. Smirnov, and V. Yu.\ Kazimirov, Phys.\ Rev.\ B \textbf{88}, 024301 (2013).
\bibitem{Ivanov-PRB2013} V. G. Ivanov, M. V. Abrashev, N. D. Todorov, V. Tomov, R. P. Nikolova, A. P. Litvinchuk, and M. N. Iliev, Phys.\ Rev.\ B \textbf{88}, 094301 (2013).
\bibitem{Yee-JKPS2003} K. J. Yee, I. H. Lee, K. G. Lee, E. Oh, D. S. Kim, and Y. S. Lim, 
    J. Kor.\ Phys.\ Soc.\ \textbf{42}, S157 (2003).
\bibitem{Bardeen-JCPA1998} C. J. Bardeen, Q. Wang, and C. V. Shank, J.\ Chem.\ Phys.\ A\ \textbf{102}, 2759 (1998).
\bibitem{Misochko-APL2007} O. V. Misochko, T. Dekorsy, S. V. Andreev, V. O. Kompanets, Yu.\ A. Matveets, A. G. Stepanov, and S. V. Chekalin, 
    Appl.\ Phys.\ Lett.\ \textbf{90}, 071901 (2007).
\bibitem{Monacelli-JCPL2017} L. Monacelli, G. Batignani, G. Fumero, C. Ferrante, Sh. Mukamel, and T. Scopigno, J.\ Phys.\ Chem.\ Lett.\ \textbf{8}, 966 (2017).
\bibitem{Tomov-JPhysCS2016} V. Tomov, P. M. Rafailov, and L. Yankova, 
    J. Phys.: Conf.\ Ser.\ \textbf{682}, 012028 (2016).
\bibitem{Boehm-JMMM2002} M. Boehm, S. Martynov, B. Roessli, G. Petrakovskii, and J. Kulda, J.\ Magn.\ Magn.\ Mater.\ \textbf{250}, 313 (2002).
\bibitem{Martynov-JMMM2006} S. Martynov, G. Petrakovskii, M. Boehm, B. Roessli, and J. Kulda, J.\ Magn.\ Magn.\ Mater.\ \textbf{299}, 75 (2006).
\bibitem{Boehm-PRB2003} M. Boehm, B. Roessli, J. Schefer, A. S. Wills, B. Ouladdiaf, E. Leli\`{e}vre-Berna, U. Staub, and G. A. Petrakovskii, Phys.\ Rev.\ B \textbf{68}, 024405 (2003).
\bibitem{Pankrats-JETPLett2003} A. I. Pankrats, G. A. Petrakovskii, M. A. Popov, K. A. Sablina, L. A. Prozorova, S. S. Sosin, G. Szimczak, R. Szimczak, and M. Baran, Pis$'$ma Zh.\ Eksp.\ Teor.\ Fiz.\ \textbf{78}, 1058 (2003) [J. Exp.\ Theor.\ Phys.\ Lett.\ \textbf{78}, 569 (2003)].
\bibitem{Fiebig-JAP2003} M. Fiebig, I. S\"{a}nger, and R. V. Pisarev, 
J.\ Appl.\ Phys.\ \textbf{93}, 6960 (2003).
\bibitem{Petrova-JETP2018} A. E. Petrova and A. I. Pankrats, Zh.\ Eksp.\ Teor.\ Fiz.\ \textbf{153}, 615 (2018).
\bibitem{Lovesey-JPCM2009-1} S. W. Lovesey and U. Staub, J. Phys.: Condens.\ Matter \textbf{21}, 142201 (2009).
\bibitem{Arima-JPCM2009} T. Arima and M. Saito, J. Phys.: Condens.\ Matter \textbf{21}, 498001 (2009).
\bibitem{Lovesey-JPCM2009-2} S. W. Lovesey and U. Staub, J. Phys.: Condens.\ Matter \textbf{21}, 498002 (2009).
\bibitem{Toyoda-PRL2015} S. Toyoda, N. Abe, S. Kimura, Y. H. Matsuda, T. Nomura, A. Ikeda, S. Takeyama, and T. Arima, 
Phys.\ Rev.\ Lett.\ \textbf{115}, 267207 (2015).
\bibitem{Lovesey-PRB2016} S. W. Lovesey, Phys.\ Rev.\ B \textbf{94}, 094422 (2016).
\bibitem{Nii-JPSJ2017}Y. Nii,  R. Sasaki,  Y. Iguchi, and  Y. Onose, J. Phys. Soc. Jpn. \textbf{86}, 024707 (2017).
\bibitem{Bossini-NPhys2018} D. Bossini, K. Konishi, S. Toyoda, T. Arima, J. Yumoto, and M. Kuwata-Gonokami, Nature Phys. \textbf{14}, 370 (2018).
\bibitem{ALeksandrov-PSS2003} K. S. Aleksandrov, B. P. Sorokin, D. A. Glushkov, L. N. Bezmaternykh, S. I. Burkov, and S. V. Belushchenko, Phys.\ Sol.\ State \textbf{45}, 41 (2003).
\bibitem{Kutt-IEEE1992} W. A. K\"{u}tt, W. Albrecht, and H. Kurz, IEEE J. Quantum Electron.\ \textbf{QE-28}, 2434 (1992).
\bibitem{Min-APL1990} L. Min and R. J. Dwayne Miller, Appl. Phys. Lett. \textbf{56}, 524 (1990).
\bibitem{Dekorsy-PRL1995} T. Dekorsy, H. Auer, C. Waschke, H. J. Bakker, H. G. Roskos, H. Kurz, V. Wagner, and P. Grosse, Phys.\ Rev.\ Lett. \textbf{74}, 738 (1995).
\bibitem{Ruhman-IEEE1988} S. Ruhman, A. G. Joly, and K. A. Nelson, IEEE J. Quantum Electron.\ \textbf{24}, 460 (1988).
\bibitem{Liu-PRL1995} Y. Liu, A. Frenkel, G. A. Garrett, J. F. Whitaker, S. Fahy, C. Uher, and R. Merlin, Phys.\ Rev.\ Lett. \textbf{75}, 334 (1995).
\bibitem{fit} The fitting was done after the numerical band pass filter process was applied to the experimental data to increase the precision of the former. The passbands were 4.2--12~THz for $\phi=0^\circ$, 4--12~THz for $\phi=45^\circ$, and 6.5--13~THz for $\phi=90^\circ$.
\bibitem{Loudon-RS1963} R. Loudon, Proc.\ Royal\ Soc.\ A \textbf{275}, 218 (1963).
\bibitem{Dougherty-PRB1994} Th. P. Dougherty, G. P. Wiederrecht, K. A. Nelson, M. H. Garrett, H. P. Jenssen, and C. Warde, Phys.\ Rev.\ B\ \textbf{50}, 8996 (1994).
\bibitem{Pisarev-PRB2016} R. V. Pisarev, M. A. Prosnikov, V. Yu. Davydov, A. N. Smirnov, E. M. Roginskii, K. N. Boldyrev, A. D. Molchanova, M. N. Popova, M. B. Smirnov, and V. Yu. Kazimirov, Phys.\ Rev.\ B\ \textbf{93}, 134306 (2016).
\bibitem{Korenev-Nphys2016} V. L. Korenev, M. Salewski, I. A. Akimov, V. F. Sapega, L. Langer, I. V. Kalitukha, J. Debus, R. I. Dzhioev, D. R. Yakovlev, D. M\"{u}ller, C. Schr\"{o}der, H. H\"{o}vel, G. Karczewski, M. Wiater, T. Wojtowicz, Yu.\ G. Kusrayev, and M. Bayer, 
Nature Phys.\ \textbf{12}, 85 (2016).
\bibitem{Nova-NPhys2016} T. F. Nova, A. Cartella, A. Cantaluppi, M. F\"{o}rst, D. Bossini, R. V. Mikhaylovskiy, A. V. Kimel, R. Merlin, and A. Cavalleri, 
Nature Phys.\ \textbf{13}, 132 (2017).
\bibitem{Azzam} R.\ M.\ A.\ Azzam and N.\ M.\ Bashara, \textit{Ellipsometry and polarized light} (North-Holland Publishing Company, 1977).
\end{thebibliography}
\end{document}